\documentclass[aps,prl,reprint,amsmath,amssymb,groupedaddress,notitlepage]{revtex4-1}
\usepackage[utf8]{inputenc}
\usepackage{bm}
\usepackage{enumerate}
\usepackage{graphicx}
\usepackage{xcolor}
\usepackage{tikz}
\usetikzlibrary{arrows}
\usepackage{hyperref}
\hypersetup{colorlinks=true,
linkcolor=blue,
citecolor=blue,
urlcolor=blue,
filecolor=blue,
hypertexnames=false
}

\newcommand{\ua}{\uparrow}
\newcommand{\da}{\downarrow}

\def\bra#1{\langle{#1}|}
\def\ket#1{|{#1}\rangle}

\begin{document}

\begin{abstract}
We propose a feasible and scalable quantum-dot-based implementation of a singlet-only spin qubit which is to leading order intrinsically insensitive to random effective magnetic fields set up by fluctuating nuclear spins in the host semiconductor.
Our proposal thus removes
an important
obstacle for further improvement of spin qubits hosted in high-quality III-V semiconductors such as GaAs.
We show how the resulting qubit could be initialized, manipulated, and read out by electrical means only, in a way very similar to a triple-dot exchange-only spin qubit.
Due to the intrinsic elimination of the effective nuclear fields from the qubit Hamiltonian, we find an improvement of the dephasing time $T_2^*$ of several orders of magnitude as compared to similar existing spin qubits.
\end{abstract}

\title{The exchange-only singlet-only spin qubit}

\author{Arnau Sala}
\affiliation{Department of Physics, NTNU, Norwegian University of Science and Technology, 7491 Trondheim, Norway}

\author{Jeroen Danon}
\affiliation{Department of Physics, NTNU, Norwegian University of Science and Technology, 7491 Trondheim, Norway}

\date{\today}

\maketitle

Spin qubits in semiconductor quantum dots are one of the more promising scalable qubit implementations put forward so far~\cite{RevQD}. 
The original proposal almost two decades ago~\cite{LossDV} was rapidly followed by early experimental successes, including demonstration of the principles of qubit initialization, manipulation, and read-out~\cite{Koppens,Elzerman}.
At the same time, two main challenges for further progress were identified:
(i) Single-qubit manipulation requires highly localized oscillating magnetic fields, which are very hard to realize in practice.
(ii) All high-quality III-V semiconductors (such as GaAs) consist of atoms carrying non-zero nuclear spin, and the fluctuating ensemble of nuclear spins in each quantum dot couples to the spin of localized electrons through hyperfine interaction.
This coupling causes spin relaxation~\cite{PhysRevB.66.155327} and yields random effective local magnetic fields acting on the electron spins, which present an important source of qubit decoherence~\cite{Merkulov,klg}.
Most of the work in the field of semiconductor spin qubits in the past decades has been aimed at overcoming these two challenges.

One proposed way to overcome the requirement of oscillating magnetic fields is to
use a material with relatively strong spin-orbit interaction (such as InAs), in which coherent spin rotations could be achieved by the application of oscillating \emph{electric} fields~\cite{Golovach,flindt:prl,Nadj}.
A drawback is that spin-orbit interaction contributes to qubit relaxation~\cite{Khaetskii2001} and also interferes with the spin-to-charge conversion commonly used for qubit initialization and read-out~\cite{PhysRevB.80.041301}.
Another approach is to
encode the qubit in a \emph{multi}-electron spin state, which enables qubit control through (gate-tunable) exchange interactions~\cite{DiVincenzo2000}:
Using two-electron spin states in a double quantum dot, one can define a qubit in the unpolarized singlet-triplet ($S$-$T_0$) subspace, which allows for electrical control of qubit rotations along one axis of the Bloch sphere~\cite{Petta,Taylor2007};
and recently it was realized that with one more quantum dot (and electron) one can use two three-electron spin states to define a qubit that has \emph{two} of such control axes~\cite{Medford}.
The resulting triple-dot exchange-only (XO) qubit can thus be fully operated by electrical means only~\cite{PhysRevB.82.075403,Medford,Russ}.
The downside of using exchange-operated spin qubits is their increased sensitivity to charge noise, either coming from environmental charge fluctuations or directly from the gates.
However, recent work indicates that symmetric operation of such qubits could greatly reduce their sensitivity to charge noise~\cite{Martins2016,Reed2016,Shim2016}.

These successes thus eliminated the need for highly localized oscillating magnetic fields, leaving the problem of the nuclear spins as the main intrinsic obstacle for further progress~\cite{Coish2005,Medford2,Mehl2013,Hung2014}.
Common approaches to overcome this problem include devising hyperfine-induced feedback cycles, where driving the electronic spins out of equilibrium results in a suppression of the fluctuations of the nuclear spin ensemble~\cite{rudnercooling,Vink,Rudner2011,Frolov,Bluhm},
as well as optimizing complex echo pulsing schemes, where the dominating frequencies in the spectrum of the nuclear spin fluctuations are effectively filtered out~\cite{Bluhm2010a,Malinowski2016,Rohling2016}, 
or via Hamiltonian parameter estimation to operate the qubit with precise knowledge of the environment~\cite{Shulman2014}.
Although some of these ideas led to significantly prolonged coherence times, they all involve a large cost in overhead for qubit operation.
Another promising approach is to host spin qubits in isotopically purified silicon, which can be (nearly) nuclear-spin-free~\cite{Zwanenburg,Eng,Veldhorst2014}, but the stronger charge noise and the extra valley degree of freedom complicate their operation.

Here we propose a new type of spin qubit that can be hosted in GaAs-based quantum dots, but (i) is intrinsically insensitive to the nuclear fields in the dots and (ii) can be operated fully electrically, similar to the triple-dot XO qubit.
The idea is to encode the qubit in a \emph{singlet-only} subspace, which is known to be ``decoherence-free'' for spin qubits (in the sense that fluctuating Zeeman fields do not act inside the subspace)~\cite{Bacon2000,Lidar2003}.
It turns out that a system of four spin-$\tfrac{1}{2}$ particles hosts such a subspace~\cite{Scarola2004,Antonio2013}:
Among the 16 different four-particle spin states there are two singlets, thus providing a decoherence-free two-level subspace.
Below we present a feasible implementation of a qubit in this subspace, using four electrons in a quadruple quantum dot.
We include a clearly outlined scheme for initialization, manipulation and read-out of this qubit, as well as an investigation of its performance in realistic circumstances.
We find that, at the price of a slight increase in complexity beyond the triple-dot XO setup, our qubit has superior coherence properties, extending $T_2^*$ by orders of magnitude, while still having a highly scalable design.

%
%

\begin{figure}
	\includegraphics[width=\linewidth]{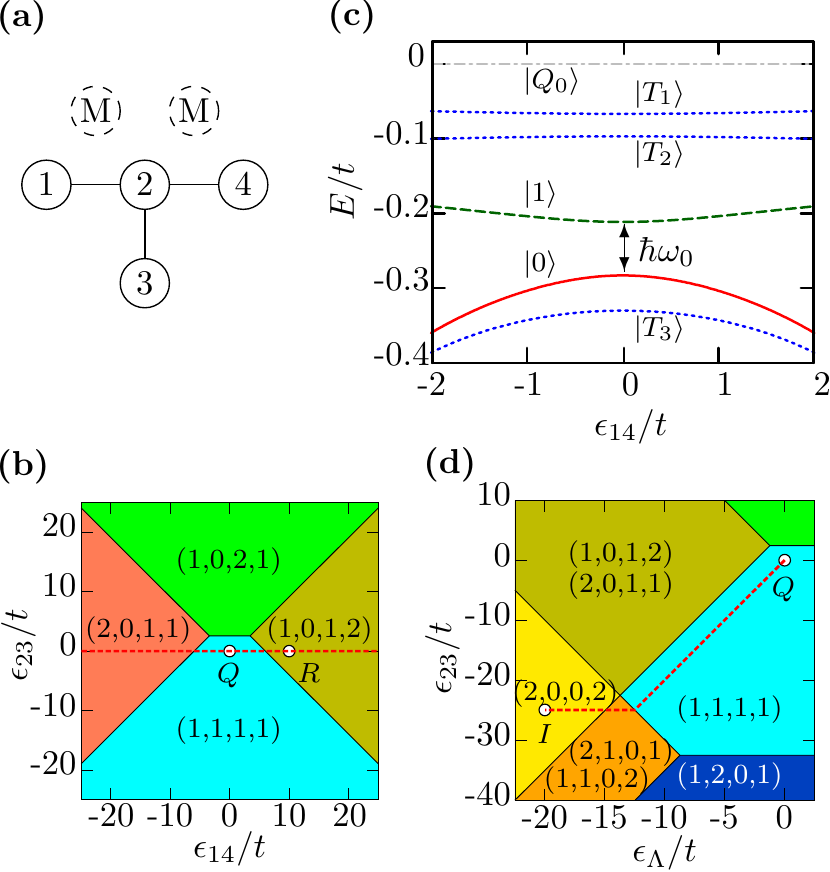}
	\caption{(color online). (a) Schematic representation of the quadruple-dot geometry. The four dots are labeled 1--4 and solid lines indicate which dots are tunnel coupled. The dashed circles labeled `M' show suggested positions for charge sensors.
		(b) Charge stability diagram in the four-electron regime as a function of $\epsilon_{14}$ and $\epsilon_{23}$, using $U=50\,t$, $U_c=15\,t$ and $\epsilon_\Lambda=0$. The red dashed line shows the tuning axis along which the qubit is operated.
		(c) Spectrum of the subspace with $S_z=0$ along the red dashed line in (b), with $t_{12}=t_{24}=\frac{4}{3}t_{23}= t$. Only the lowest part of the spectrum is shown. The red solid and green dashed lines correspond to the two singlet states that form the qubit.
		(d) Charge stability diagram as a function of $\epsilon_{\Lambda}$ and $\epsilon_{23}$ with $\epsilon_{14}=0$ and further the same parameters as in (b). The red dashed line shows the path we suggest for qubit initialization.\label{fig:first}}
\end{figure}

\textit{The qubit.}---We propose a setup in which four quantum dots are arranged in a T-like geometry, as shown in Fig.~\ref{fig:first}(a), where solid lines connect dots that are tunnel coupled.
Nearby charge sensors, indicated by `M', can be used to monitor the charge state $(N_1,N_2,N_3,N_4)$ of the quadruple dot, where $N_i$ is the number of excess electrons on dot number $i$, as labeled in Fig.~\ref{fig:first}~\cite{Note1}.
To describe this system, we use a Hubbard-like Hamiltonian~\cite{burkard_prb,Taylor},
\begin{align}
\hat{H} = {} & {} \sum_{i} \left[ \frac{U}{2}\, \hat{n}_{i} (\hat{n}_{i} -1) - V_i\, \hat{n}_{i} \right]
+ \sum_{\langle i,j\rangle} U_c\, \hat{n}_{i} \hat{n}_{j}
\notag \\ {} & {}
- \sum_{\langle i,j \rangle, \alpha} \frac{t_{ij}}{\sqrt{2}}\, \hat{c}^\dagger_{i,\alpha} \hat{c}_{j,\alpha}
+ \sum_{i,\alpha} \frac{E_\text{Z}}{2}\,\hat{c}^\dagger_{i,\alpha} \sigma^z_{\alpha \alpha} \hat{c}_{i,\alpha}, \label{eq:ham}
\end{align}
where $\hat{n}_i = \sum_{\alpha} \hat c_{i,\alpha}^\dagger \hat c_{i,\alpha}$ with $\hat c_{i,\alpha}^\dagger$ the creation operator for an electron with spin $\alpha$ on dot $i$.
The first line of Eq.~(\ref{eq:ham}) describes the electrostatic energy of the system:
The first term accounts for the on-site Coulomb interaction of two electrons occupying the same dot, the second term adds a local offset of the potential energy that can be controlled via gating, and the last term describes the cross-capacitance between neighboring dots.
To this we added (spin-conserving) tunnel couplings between neighboring dots, characterized by coupling energies $t_{ij}$, and a uniform Zeeman splitting of the electronic spin states induced by an external magnetic field applied along the $z$-direction, where $E_\text{Z} = g\mu_\text{B} B$ is the Zeeman energy, with $g$ the effective $g$-factor ($g \approx -0.4$ in GaAs), $\mu_\text{B}$ the Bohr magneton, and $B$ the magnitude of the applied field.

We can use the electrostatic part of $\hat H$ to find the charge ground state as a function of the gate-induced offsets $V_i$.
For convenience, we introduce the tuning parameters $\epsilon_{14}=(V_4-V_1)/2$, $\epsilon_{23}=(V_3-V_2)/2$, $\epsilon_\Lambda = (-V_1+V_2 +V_3 - V_4)/4$, and $\epsilon_\Sigma=(V_1+V_2+V_3+V_4)/4$, where we fix $\epsilon_\Sigma = \tfrac{3}{4}U_c$.
Focusing on the four-electron regime, we show a part of the resulting charge stability diagram as a function of $\epsilon_{14}$ and $\epsilon_{23}$ in Fig.~\ref{fig:first}(b), where we have set $\epsilon_\Lambda = 0$, $U=50\, t$, and $U_c=15\, t$ ($t$ being our unit of energy, of the order of the tunnel coupling energies).
Our region of interest is the `top' of the (1,1,1,1) charge region, where exchange effects due to the vicinity of the (2,0,1,1), (1,0,2,1), and (1,0,1,2) charge regions can be significant and are effectively tunable through the gate potentials $V_i$.
We note here that we will assume throughout that the orbital level splitting on the dots is the largest energy scale in the system (larger than $U$), so we will only include states involving double occupation ($N_i=2$) if the two electrons are in a singlet state.

We now include finite tunnel coupling energies $t_{ij}$ and a Zeeman energy $E_\text{Z}$, and investigate the spectrum of $\hat H$ in more detail.
The red dashed line in Fig.~\ref{fig:first}(b) indicates where $\epsilon_{23}=0$, and along this line $\epsilon_{14}$ parametrizes a `linear tilt' of the potential of the three dots 1, 2, and 4, equivalent to the triple-dot detuning parameter that is used to operate the XO qubit (see Refs.~\cite{Taylor,Medford,Medford2}).
In Fig.~\ref{fig:first}(c) we plot the resulting spectrum of the six lowest-lying states with $S_z=0$ along this line, as a function of $\epsilon_{14}$, where we have set $t_{12}=t_{24}=\frac{4}{3}t_{23}= t$ and $E_\text{Z} = 1.875\, t$.
In the plot we can identify one quintuplet state $\ket{Q_0}$ (gray dotted-dashed), three triplet states $\ket{T_{1,2,3}}$ (blue dotted), and two singlets (green dashed and red solid).
The ten other spin states, having $S_z=\pm 1, \pm 2$, are split off by multiples of $E_\text{Z}$ and not shown in the plot.

The two singlets we propose to use as qubit basis states are marked $\ket{0}$ and $\ket{1}$ in Fig.~\ref{fig:first}(c) and read to lowest (zeroth) order in the tunnel couplings $t_{ij}$
\begin{align} \label{eq:states}
|1\rangle = {} & {}   \left| S_{14}  S_{23} \right\rangle, \\
|0\rangle = {} & {}   \frac{1}{\sqrt{3}} \left\{ \left| S_{13} S_{24} \right\rangle + \left| S_{12} S_{34} \right\rangle \right\},
\end{align}
where $S_{ij}$ denotes a singlet pairing of the two electrons in dots $i$ and $j$.
As an example: one can write explicitly $\ket{1} = \{ \ket{\ua\ua\da\da} - \ket{\ua\da\ua\da} - \ket{\da\ua\da\ua} +\ket{\da\da\ua\ua} \}/2$.

Close to the central point $\epsilon_{14}=0$, marked $Q$ in Fig.~\ref{fig:first}(b), exchange effects are small in $t/\Delta$, where $\Delta = U - 3U_c$ is the half-width of the (1,1,1,1) charge region along the detuning axis $\epsilon_{14}$, and we thus treat the tunnel couplings as perturbations.
Including only the nearby charge states (2,0,1,1), (1,0,2,1), and (1,0,1,2), we can project $\hat H$ to the qubit subspace spanned by $\ket{0}$ and $\ket{1}$, yielding to second order in the $t_{ij}$
\begin{align}\label{eq:eff}
\hat H_{\rm qb} = 
\frac{1}{4}(J_{12}+J_{24}-2J_{23})\hat\sigma^z
+\frac{\sqrt 3}{4}(J_{12} - J_{24})\hat\sigma^x,
\end{align}
where we subtracted a constant offset.
The $\hat \sigma^{x,z}$ denote Pauli matrices, and the relative magnitudes of the exchange energies $J_{12} = t^2_{12}/(\Delta+\epsilon_{14})$, $J_{24} = t^2_{24}/(\Delta-\epsilon_{14})$, and $J_{23} = t^2_{23}/\Delta$, can be controlled by the detuning parameter $\epsilon_{14}$ (note that we have set $\epsilon_{23} = 0$).
We make two observations:
(i) The qubit splitting at zero detuning, $\hbar\omega_0 = (t_{12}^2 + t_{24}^2 - 2t_{23}^2)/2\Delta$, vanishes if all three tunnel couplings are equal; ideally, one tunes $t_{12}= t_{24} \neq t_{23}$.
(ii) The structure of this Hamiltonian is fully equivalent to that of the triple-dot XO qubit [cf.~Eq.~(5) in Ref.~\cite{Taylor}], including its qualitative dependence on the detuning parameter. Thus, our qubit can be operated analogously to the XO qubit, i.e., by static pulsing \cite{Medford} or resonant driving \cite{Medford2}, and the point $Q$ is a sweet spot where the qubit is to lowest order insensitive to noise in $\epsilon_{14}$.

\textit{Qubit operation.}---Qubit rotations are most conveniently achieved using resonantly driven Rabi oscillations~\cite{Medford2}.
For small detuning, $|\epsilon_{14}| \ll \Delta$, we can expand $\hat H_\text{qb}$ to linear order in $\epsilon_{14}$, yielding
\begin{align} \label{eq:heff}
\hat{H}_\text{qb} = \frac{1}{2}\hbar \omega_0 \hat{\sigma}^z - \frac{\sqrt 3}{2} \frac{t^2 \epsilon_{14}}{\Delta^2} \hat{\sigma}^x,
\end{align}
where we used $t_{12} = t_{24} = t$.
A harmonic modulation of the detuning, $\epsilon_{14} = A\, \cos(\omega \tau)$, will thus induce Rabi rotations of the qubit which will have at the resonance condition $\omega = \omega_0$ a Rabi period of $T_\text{Rabi} = 4\pi\hbar\Delta^2/(\sqrt 3 t^2 A)$.
Using again $\Delta = 5\, t$ (consistent with the realistic parameters $t = 20~\mu$eV, $U=1$~meV, and $U_c=0.3$~meV), a moderate driving amplitude of $A = 2.5~\mu$eV would yield a rotation time $T_\text{Rabi} \approx 50$~ns.

Read-out of the qubit can be performed by spin-to-charge conversion, in a similar way as in the double-dot $S$-$T_0$ \cite{Petta} and triple-dot XO~\cite{Medford,Medford2} qubits.
The detuning $\epsilon_{14}$ is quickly pulsed to the point marked $R$ in Fig.~\ref{fig:first}(c), which lies in the (1,0,1,2) charge region.
There are only two accessible (1,0,1,2)-states with $S_z=0$:
The two electrons on dot 4 must be in a singlet state, but the electrons on dot 1 and 3 can form either a singlet $S$ or unpolarized triplet $T_0$.
Only the singlet-singlet configuration couples adiabatically to one of the qubit states (the state $\ket{0}$).
After pulsing to $R$, the qubit state $\ket{0}$ will thus transition to a (1,0,1,2) charge configuration whereas the state $\ket{1}$ remains in a spin-blockaded (1,1,1,1) state.
Subsequent charge sensing amounts to a projective measurement of the qubit state.
One requirement is that the detuning pulse has to be fast enough so that spin-flip transitions from $\ket{0}$ to one of the lower-lying states with $S_z = 1,2$, which are crossed at $\epsilon_{14} \sim E_\text{Z}$, are very unlikely.
(Note that exactly the same condition holds for the triple-dot XO qubit measurement scheme, where the spin-$\tfrac{1}{2}$ state connected to $\ket{0}$ crosses a spin-$\tfrac{3}{2}$ state~\cite{Medford2}.)

Initialization of the qubit can be achieved in a similar way.
The simplest procedure is to pulse to a point in gate space where there is one unique singlet-only ground state, such as the point marked $I$ in the (2,0,0,2) charge region, see Fig.~\ref{fig:first}(d).
After waiting long enough, the system will have relaxed to this ground state, and a fast pulse back to the qubit tuning $Q$ will yield a qubit prepared in $\ket{0}$.
The path we propose for this pulse is marked in Fig.~\ref{fig:first}(d) by a red dashed line: 
first $\epsilon_\Lambda$ is increased until the edge of the (1,1,1,1) charge region is reached, after which both $\epsilon_\Lambda$ and $\epsilon_{23}$ are increased simultaneously until the system reaches the point $Q$.
For this pulse the same condition holds as for the read-out pulse: it should be fast enough to not allow for spin-flip transitions into the lower-lying states with $S_z=1,2$~\cite{Note2}.

\textit{Decoherence.}---The main source of decoherence in GaAs-based spin qubits is known to be the fluctuating bath of nuclear spins that couples to the qubit states through hyperfine interaction~\cite{Merkulov,RevQD,Medford2}.
The effect of the ensemble of $\sim 10^6$ nuclear spins in each quantum dot can to good approximation be modeled as a randomly and slowly fluctuating effective magnetic field ${\bf K}_i$ acting on the electrons localized in the dot $i$.
The fluctuations are slow enough that the field can be considered as static on the time scale of a single qubit operation, but it varies randomly over the course of many measurement cycles.
The r.m.s.\ value of these random fields was reported to be $K = 1$--3~mT in typical GaAs quantum dots~\cite{Johnson,Koppens,Medford2}.
The resulting uncertainty in the qubit level splitting translates to a decoherence time $T_2^*$ of tens of ns, and forms at present the bottleneck for further improvement of the performance of GaAs-based spin qubits.

To understand the effect of hyperfine interaction on the singlet-only qubit, we write the effective Hamiltonian
\begin{align} \label{eq:hhf0}
\hat{H}_\text{hf} = \frac{g\mu_\text{B}}{2} \sum_{i,\alpha,\beta}  \hat{c}^\dagger_{i,\alpha} \mathbf{K}_i \cdot \boldsymbol{\sigma}_{\alpha \beta} \hat{c}_{i,\beta}.
\end{align}
and project this Hamiltonian to the qubit subspace,
\begin{align} \label{eq:hhfqb}
	\hat{H}_\text{hf,qb} = 0.
\end{align}
This confirms that, to leading order, the nuclear fields do not affect the qubit and thus do not cause any decoherence.
The hyperfine Hamiltonian does, however, couple both qubit states to all nine four-electron triplet states.
Coupling to the triplet states with $S_z=\pm 1$ is mediated by $K_{i}^x$ and $K_{i}^y$, but transitions to these states are strongly suppressed by the large Zeeman energy $E_\text{Z}$.
The $z$-components of the nuclear fields couple $\ket{0}$ and $\ket{1}$ to $\ket{T_{1,2,3}}$, and this coupling (i) can cause leakage out of the qubit space, analogous to leakage to the spin-$\tfrac{1}{2}$ quadruplet state in the triple-dot XO qubit, and (ii) can yield a higher-order shift in the qubit splitting, contributing to qubit decoherence.
Both effects are suppressed by the small factor $g\mu_\text{B}K/J$ (where $J$ is the typical energy scale of the exchange energies $J_{ij}$), and the decoherence time resulting from the fluctuations of the qubit splitting~\cite{Note3} can be estimated as $T_2^*\sim \hbar J / (g\mu_\text{B}K)^2$.
For typical parameters ($J = 2~\mu$eV and $K = 1$~mT) this would present an improvement of two orders of magnitude over other GaAs-based spin qubits, where $T_2^*\sim \hbar / g\mu_\text{B}K$.

\begin{figure}
	\includegraphics[scale=1]{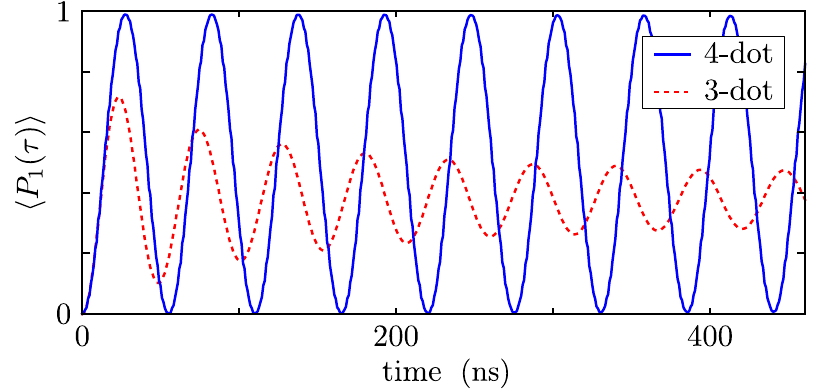}
	\caption{(color online). Solid blue: Calculated time-dependent expectation value of $\ket{1}\bra{1}$ after initializing in $|0\rangle$ and driving resonantly with $\epsilon_{14} \propto \cos(\omega_0\tau)$, averaged over 2500 random configurations of the nuclear fields.
		Dashed red: Equivalent result for the triple-dot XO setup, using the same parameters. \label{fig:third}}
\end{figure}

To support these claims, we perform numerical simulations of resonant driving of the qubit.
We project the Hamiltonian (\ref{eq:ham}) to the 12-dimensional subspace of all (1,1,1,1), (2,0,1,1), (1,0,2,1), and (1,0,1,2) states with $S_z = 0$.
We diagonalize the resulting Hamiltonian at the point $Q$, see Fig.~\ref{fig:first}(b), using the same parameters as before and specifying $t = 16~\mu$eV; this yields all eigenstates at $\epsilon_{14} = 0$ as well as the qubit splitting $\hbar\omega_0$.
We initialize in the lowest-lying singlet state $\ket{0}$, and then let the system evolve under the Hamiltonian $\hat H + \hat H_\text{hf}$ where we include resonant driving $\epsilon_{14} = A\,\cos(\omega_0\tau)$ and four random nuclear fields ${\bf K}_i$.
In Fig.\ \ref{fig:third} (solid blue) we show the resulting time-dependent probability to find the system in $\ket{1}$, where we used $A = 2.5~\mu$eV and averaged over 2500 random nuclear field configurations with the $g\mu_\text{B}K_i^{x,y,z}$ drawn from a normal distribution with mean zero and $\sigma = 0.07~\mu$eV.
We observe eight Rabi oscillations in $\sim 450$~ns without any significant decay \cite{Note4}.
Of course, at longer times eventually leakage out of the qubit space as well as higher-order corrections due to the nuclear fields will suppress the oscillations in $\langle P_1(\tau)\rangle$.

As a comparison, we also performed equivalent simulations of resonant driving of a triple-dot XO qubit (cf.~Ref.~\cite{Medford2}), using exactly the same parameters (basically setting $t_{23} = 0$ and adjusting $\omega_0$ to the new qubit splitting).
The result is plotted with a red dashed line in Fig.~\ref{fig:third}; in this case the hyperfine-induced decay of $\langle P_1(\tau)\rangle$ is already significant in the first few Rabi periods.
The clear contrast between the two curves illustrates the improvement presented by our quadruple-dot XO qubit.

\textit{Relaxation.}---Electron-phonon coupling can contribute to qubit relaxation, i.e., induce dissipative transitions from $\ket{1}$ to $\ket{0}$.
The associated relaxation rate can be estimated  using Fermi's golden rule, $\Gamma_\text{rel}=\frac{2\pi}{\hbar} \sum_f | \langle f | \hat{H}_\text{e-ph} | i \rangle |^2 \delta(E_f-E_i)$, where the initial state is $\ket{1}\ket{\text{vac}}$ (with $\ket{\text{vac}}$ denoting the phonon vacuum) and the sum runs over all possible final states $\ket{0}\ket{1_{{\bf k},p}}$ where one phonon has been created with wave vector ${\bf k}$ and polarization $p$.
We use an electron-phonon Hamiltonian
\begin{align}
\hat{H}_\text{e-ph} = \sum_{\mathbf{k},p} \lambda_{\mathbf{k},p} \hat{\rho}_{\mathbf{k}} ( \hat{a}_{\mathbf{k},p} + \hat{a}^\dagger_{-\mathbf{k},p} ),
\end{align}
where $\hat{a}^\dagger_{\mathbf{k},p}$ creates a phonon in mode $\{\mathbf{k},p\}$, $\hat{\rho}_{\mathbf{k}}$ is the Fourier transform of the electronic density matrix, and $\lambda_{\mathbf{k},p}$ are the coupling parameters (see e.g., Ref.~\cite{PhysRevB.74.045320}.
At typical qubit splittings the coupling to piezoelectric phonons dominates, in which case an explicit evaluation of $\Gamma_\text{rel}$ yields to leading order in $t/\Delta$ the estimate~\cite{Note2}
\begin{align}
\Gamma_\text{rel} \approx
  \frac{t^4}{\Delta^4} \frac{\omega_0^3 d^3}{v^3} \frac{(eh_{14})^2}{10\pi\hbar\rho v^2 d},
\end{align}
where $v$ is the phonon velocity (for convenience now assumed equal for all three polarizations), $d$ the distance between neighboring dots, $h_{14}$ the piezoelectric constant, and $\rho$ denotes the mass density of the semiconductor (for GaAs $v \sim 4000~$m/s, $h_{14} \approx 1.45 \times 10^9$~V/m, and $\rho \approx 5300$~kg/m$^3$; see Ref.~\cite{Madelung2004}).
Setting $d = 100$~nm, this yields $\Gamma_\text{rel} = \omega_0^3 (t/\Delta)^4 (3\times 10^{-23}\,{\rm Hz}^{-2})$, which for $\Delta = 5\, t$ and $\hbar\omega_0 = 1.5~\mu$eV gives $\Gamma_\text{rel} = 0.57$~kHz.

Relaxation processes to $\ket{T_3}$ require a change of the spin state of the electrons~\cite{Note5} and are estimated to be smaller by a factor $\sim (g\mu_\text{B}K/J)^2$.
Dissipative transitions to the lower-lying states with $S_z=1,2$, require a spin-flip and are suppressed by the large Zeeman energy $E_\text{Z}$.

\textit{Conclusions.}---We propose a new, quantum-dot-based singlet-only spin qubit which is to leading order intrinsically insensitive to randomly fluctuating nuclear fields.
Our proposal thus removes the main obstacle for further improvement of spin qubits hosted in semiconductors with spinful nuclei, such as GaAs.
Its scalability, full electrical control and large coherence time make the singlet-only spin qubit one of unprecedented quality.

\textit{Acknowledgments.}---We thank M.S.~Rudner for several very helpful discussions.

%


\clearpage
\onecolumngrid
\begin{center}
\textbf{\large The exchange-only singlet-only spin qubit: Supplementary Material}
\end{center}

\setcounter{equation}{0}
\setcounter{figure}{0}
\setcounter{table}{0}
\setcounter{page}{1}
\makeatletter

\allowdisplaybreaks

\newcommand{\vk}{\mathbf{k}}
\renewcommand{\vr}{\mathbf{r}}

\renewcommand{\theequation}{S\arabic{equation}}
\renewcommand{\thefigure}{S\arabic{figure}}
\renewcommand{\thetable}{S\arabic{table}}

\vspace{1cm}
In this supplementary material we complement the results presented in the main text with several more detailed discussions.
We included
(i) a more detailed discussion of the structure of the $S^z = 0$ (1,1,1,1) subspace, including a derivation of the qubit Hamiltonian (4);
(ii) the hyperfine Hamiltonian (6) projected to the full $S^z = 0$ subspace and an investigation of its higher-order effects on the qubit;
(iii) the derivation of the estimate for the phonon-induced relaxation rate (9);
(iv) a detailed discussion of the proposed initialization procedure of the qubit;
and (v) a sketch of a possible gate design for the qubit, meant to illustrate its feasibility.

\section{Structure of the $S^z=0$ subspace and qubit Hamiltonian}

\subsection{Four spin-$\tfrac{1}{2}$ particles}

Four spin-$\tfrac{1}{2}$ particles can be in 16 states, of which six states have total spin projection $S^z_{\rm tot}=0$.
Explicitly, one can choose to write these states as
\begin{align}
	\ket{S^\alpha} & = \frac{1}{2} \big[ \ket{\ua\ua\da\da} - \ket{\ua\da\ua\da} - \ket{\da\ua\da\ua} +\ket{\da\da\ua\ua}  \big], \label{eq:sa}\\
	\ket{S^\beta} & = \frac{1}{2\sqrt{3}} \big[ \ket{\ua\ua\da\da} + \ket{\da\da\ua\ua} + \ket{\ua\da\ua\da} +\ket{\da\ua\da\ua} - 2\ket{\ua\da\da\ua} - 2\ket{\da\ua\ua\da}  \big],\label{eq:sb}\\
	\ket{T_0^\alpha} & = \frac{1}{2\sqrt{3}} \big[ 2\ket{\ua\da\da\ua} - 2\ket{\da\ua\ua\da} - \ket{\ua\da\ua\da} + \ket{\da\ua\da\ua} + \ket{\ua\ua\da\da} - \ket{\da\da\ua\ua}  \big], \\
	\ket{T_0^\beta} & = \frac{1}{2} \big[ \ket{\ua\da\ua\da} - \ket{\da\ua\da\ua} + \ket{\ua\ua\da\da} - \ket{\da\da\ua\ua}  \big], \\
	\ket{T_0^\gamma} & = \frac{1}{\sqrt{6}} \big[ \ket{\ua\da\ua\da} + \ket{\ua\da\da\ua} - \ket{\da\ua\ua\da} - \ket{\da\ua\da\ua} - \ket{\ua\ua\da\da} + \ket{\da\da\ua\ua}  \big], \\
	\ket{Q_0} & = \frac{1}{\sqrt{6}} \big[ \ket{\ua\ua\da\da} + \ket{\ua\da\ua\da} + \ket{\da\ua\ua\da} + \ket{\ua\da\da\ua} + \ket{\da\ua\da\ua} + \ket{\da\da\ua\ua}  \big].\label{eq:q0}
\end{align}
There are two singlet states, labeled $S$, three unpolarized triplet states, $T$, and one unpolarized quintuplet state $Q$.

One can write all of these states as a linear combination of two-particle product states, where each ``particle'' now represents an actual two-particle spin singlet or triplet.
There are of course three different ways to choose which pairs of spin-$\tfrac{1}{2}$ particles we treat as spin-0 or spin-1 particles, and the extensive list of possible notations reads as follows:
\begin{align}
	\ket{S^\alpha} & = \frac{1}{2} \big[ \ket{T_+^{(12)}T_-^{(34)}} + \ket{T_-^{(12)}T_+^{(34)}} - \ket{T_0^{(12)}T_0^{(34)}} - \ket{S^{(12)}S^{(34)}} \big],\\
	& = \frac{1}{2} \big[ - \ket{T_+^{(13)}T_-^{(24)}}  - \ket{T_-^{(13)}T_+^{(24)}}  + \ket{T_0^{(13)}T_0
		^{(24)}} + \ket{S^{(13)}S^{(24)}} \big],\\
	& = \ket{S^{(14)}  S^{(23)}},\\
    \ket{S^\beta} & =  \frac{1}{2\sqrt{3}} \big[ \ket{T_+^{(12)}T_-^{(34)}}  + \ket{T_-^{(12)}T_+^{(34)}}  - \ket{T_0^{(12)}T_0^{(34)}} + 3\ket{S^{(12)}S^{(34)}} \big],\\
	& = \frac{1}{2\sqrt{3}} \big[ \ket{T_+^{(13)}T_-^{(24)}}  + \ket{T_-^{(13)}T_+^{(24)}}  - \ket{T_0^{(13)}T_0^{(24)}} + 3\ket{S^{(13)}S^{(24)}} \big],\\
	& = \frac{1}{\sqrt{3}} \big[ - \ket{T_+^{(14)}T_-^{(23)}}  - \ket{T_-^{(14)}T_+^{(23)}}  + \ket{T_0^{(14)}T_0^{(23)}} \big],\\
	\ket{T_0^\alpha} & = \frac{1}{2\sqrt{3}} \big[ \ket{T_+^{(12)}T_-^{(34)}} - \ket{T_-^{(12)}T_+^{(34)}} + \ket{S^{(12)}T_0^{(34)}} - 3\ket{T_0^{(12)}S^{(34)}} \big],\\
	& = \frac{1}{2\sqrt{3}} \big[ - \ket{T_+^{(13)}T_-^{(24)}} + \ket{T_-^{(13)}T_+^{(24)}} + 3\ket{S^{(13)}T_0^{(24)}} - \ket{T_0^{(13)}S^{(24)}} \big],\\
	& = \frac{1}{\sqrt{3}} \big[ \ket{T_+^{(14)}T_-^{(23)}} - \ket{T_-^{(14)}T_+^{(23)}} + \ket{T_0^{(14)}S^{(23)}} \big],\\
	\ket{T_0^\beta} & = \frac{1}{2} \big[ \ket{T_+^{(12)}T_-^{(34)}} - \ket{T_-^{(12)}T_+^{(34)}} + \ket{S^{(12)}T_0^{(34)}} + \ket{T_0^{(12)}S^{(34)}} \big],\\
	& = \frac{1}{2} \big[ \ket{T_+^{(13)}T_-^{(24)}} - \ket{T_-^{(13)}T_+^{(24)}} + \ket{S^{(13)}T_0^{(24)}} + \ket{T_0^{(13)}S^{(24)}} \big],\\
	& = \ket{S^{(14)}T_0^{(23)}},\\
	\ket{T_0^\gamma} & = \frac{1}{\sqrt{6}} \big[ - \ket{T_+^{(12)}T_-^{(34)}}  + \ket{T_-^{(12)}T_+^{(34)}}  + 2 \ket{S^{(12)}T_0^{(34)}} \big],\\
	& = \frac{1}{\sqrt{6}} \big[ \ket{T_+^{(13)}T_-^{(24)}}  - \ket{T_-^{(13)}T_+^{(24)}}  - 2\ket{T_0^{(13)}S^{(24)}} \big],\\
	& = \frac{1}{\sqrt{6}} \big[ \ket{T_+^{(14)}T_-^{(23)}}  - \ket{T_-^{(14)}T_+^{(23)}}  - 2\ket{T_0^{(14)}S^{(23)}} \big],\\
	\ket{Q_0} & = \frac{1}{\sqrt{6}} \big[ \ket{T_+^{(12)}T_-^{(34)}} + \ket{T_-^{(12)}T_+^{(34)}} + 2\ket{T_0^{(12)}T_0^{(34)}} \big],\\
      & = \frac{1}{\sqrt{6}} \big[ \ket{T_+^{(13)}T_-^{(24)}} + \ket{T_-^{(13)}T_+^{(24)}} + 2\ket{T_0^{(13)}T_0^{(24)}} \big],\\
      & = \frac{1}{\sqrt{6}} \big[ \ket{T_+^{(14)}T_-^{(23)}} + \ket{T_-^{(14)}T_+^{(23)}} + 2\ket{T_0^{(14)}T_0^{(23)}} \big],
\end{align}
where the superscript numbers indicate which two spin-$\tfrac{1}{2}$ particles have been paired up in each singlet and triplet state.
Although at this point this might seem to be a pointless exercise in notational gymnastics, this overview can be convenient while deriving the exchange-induced matrix elements.

\subsection{Hamiltonian}

To describe the electrostatics of four electrons in four quantum dots, we use a Hubbard-like Hamiltonian~\cite{SUPburkard_prb,SUPTaylor},
\begin{align}\label{eq:sham0}
\hat{H}_{\rm es} = {} & {} \sum_{i} \left[ \frac{U}{2}\, \hat{n}_{i} (\hat{n}_{i} -1) - V_i\, \hat{n}_{i} \right]
+ \sum_{\langle i,j\rangle} U_c\, \hat{n}_{i} \hat{n}_{j},
\end{align}
with $\hat{n}_i = \sum_{\alpha} \hat c_{i,\alpha}^\dagger \hat c_{i,\alpha}$, where $\hat c_{i,\alpha}^\dagger$ creates an electron with spin $\alpha$ on dot $i$.
This Hamiltonian includes (i) a charging energy $U$ associated with doubly occupied dots (we assume the orbital level splitting on the dots to be so large that we need only consider states with $\langle 
\hat n_i \rangle = 0,1,2$, where $\langle 
\hat n_i \rangle = 2$ must correspond to a two-electron spin-singlet state), (ii) on-site potentials $V_i$ that can be tuned through the electrostatic gates, and (iii) a cross-capacitance $U_c$ between neighboring dots.

Different charge states, that are split in energy by $\hat H_{\rm es}$, are still highly degenerate due to the large number of allowed four-particle spin configurations:
In a (1,1,1,1) charge state the four electrons can be in any of the 16 possible spin states, of which 6 are listed above.
This degeneracy can be partly lifted by the Zeeman effect due to the application of an external magnetic field.
To describe this Zeeman splitting, we add a corresponding term to the Hamiltonian,
\begin{align}
\hat H_{\rm Z} =  \sum_{i,\alpha} \frac{E_\text{Z}}{2}\,\hat{c}^\dagger_{i,\alpha} \sigma^z_{\alpha \alpha} \hat{c}_{i,\alpha}
= E_{\rm Z} \hat S^z_{\rm tot},
\end{align}
where $\hat S^z_{\rm tot} = \sum_i \hat S^z_{i}$ is the projection operator for the total (four-particle) spin along the $z$-axis, which is chosen to be parallel to the applied field.
For finite $E_{\rm Z}$, all states with different $S^z_{\rm tot}$ will now be split in energy; states with the same $S^z_{\rm tot}$ (and the same charge configuration), such as the six states listed in (\ref{eq:sa})--(\ref{eq:q0}), remain degenerate.

Finally, we add a term that describes the finite tunnel coupling between neighboring dots and lifts the degeneracy,
\begin{align} \label{eq:shamt}
	\hat{H}_\text{t} =&  - \sum_{\langle i,j \rangle, \alpha} \frac{t_{ij}}{\sqrt{2}}\, \hat{c}^\dagger_{i,\alpha} \hat{c}_{j,\alpha},
\end{align}
where the specific geometry considered here (a T-like setup) dictates that only $t_{12}$, $t_{23}$, and $t_{24}$ are non-zero.

\subsection{Effective Hamiltonian at qubit operation point}

We start by introducing four independent tuning parameters,
\begin{align}
\epsilon_{14} {} & {} = \frac{1}{2}(V_4-V_1), \\
\epsilon_{23} {} & {} = \frac{1}{2}(V_3-V_2), \\
\epsilon_\Lambda {} & {} = \frac{1}{4}(-V_1+V_2 +V_3 - V_4), \\
\epsilon_\Sigma {} & {} =\frac{1}{4}(V_1+V_2+V_3+V_4),
\end{align}
which parametrize the set of gate-induced energy offsets $\{V_1,V_2,V_3,V_4\}$.
We define the qubit operation point $Q$ to have the tuning $\epsilon_{14} = \epsilon_{23} = \epsilon_\Lambda = 0$, corresponding to setting $V_1 = V_2 = V_3 = V_4$.
We further fix the overall energy offset $\epsilon_\Sigma = \tfrac{3}{4}U_c$ so that the (1,1,1,1) state $\ket{Q_0}$ has zero energy.

\begin{figure}[t!]
\begin{center}
\includegraphics{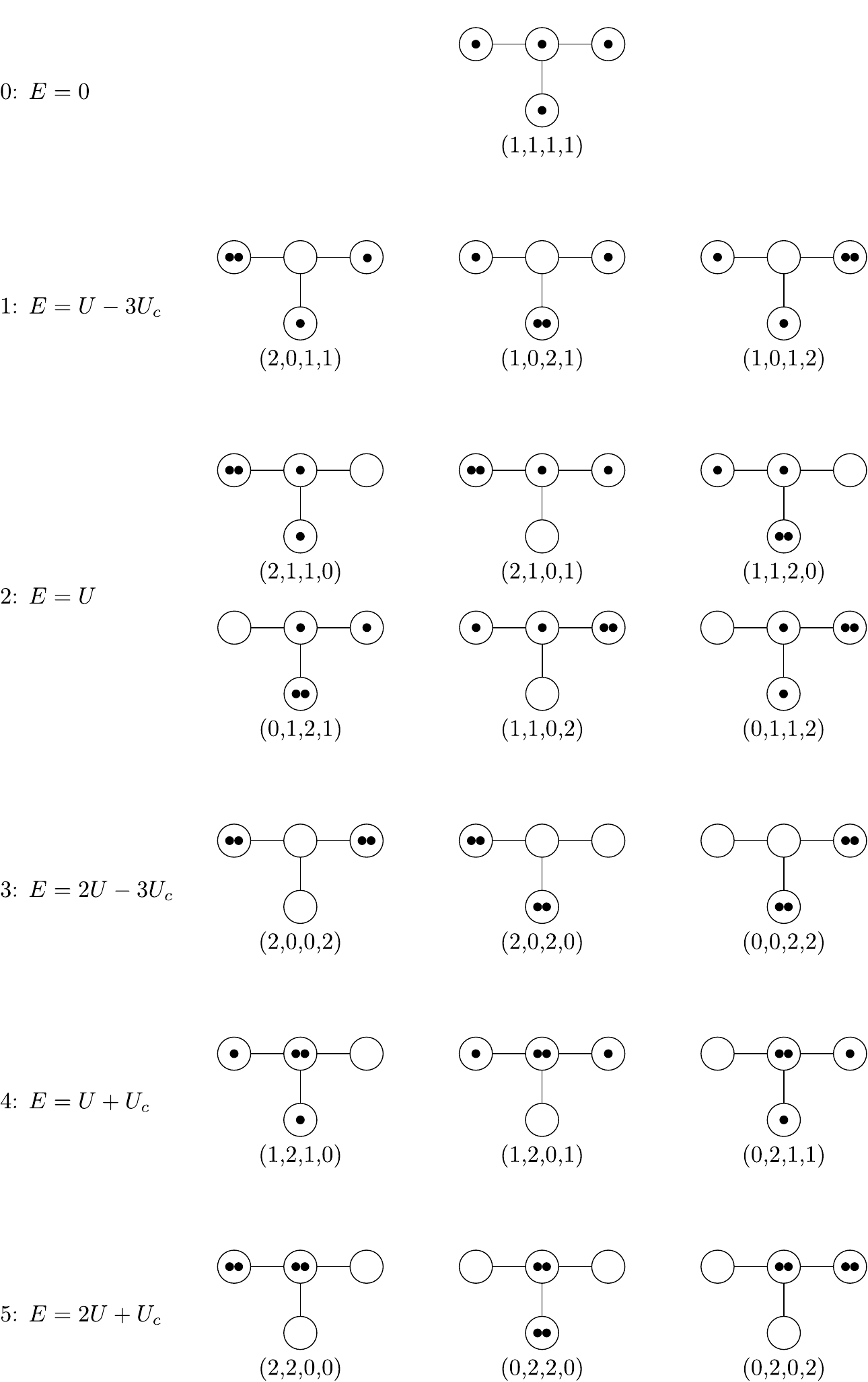}
\caption{Ground state and all excited states of $\hat H_{\rm es}$ at the qubit operation point $Q$, assuming that there are four electrons in the system and only on-site occupation numbers $n_i = 0,1,2$ are allowed. The excitation energies are given on the left side of the figure. We note that only the states in groups 1 and 4 are connected to the ground state 0 by a single tunneling event.}\label{fig:charges}
\end{center}
\end{figure}
Assuming that $U>3U_c$, the charge ground state around $Q$ is (1,1,1,1) and the lowest-lying excited states are (2,0,1,1), (1,0,2,1), and (1,0,1,2), which all have an excitation energy of $U-3U_c \equiv \Delta$, see Fig.~\ref{fig:charges}.
We assume that the typical energy scale of the tunnel couplings is much smaller than this excitation energy, $t\ll \Delta$, which allows us to evaluate all exchange effects using perturbation theory in the tunnel couplings $t_{ij}$.

To first order in the $t_{ij}$, only the charge states collected in groups 1 and 4 in Fig.~\ref{fig:charges} are coupled to the (1,1,1,1) charge state.
Assuming that $U_c \sim U$, we see that the states in group 4 are much higher in energy as compared to the ones in group 1, so to first approximation we can neglect the states in group 4 in our evaluation of the exchange corrections inside the (1,1,1,1) charge subspace.

We thus start from a 28-dimensional Hamiltonian, including all 16 allowed spin configurations of the (1,1,1,1) state and all four spin configurations allowed for any of the three charge states in group 1.
We decouple sectors in the Hamiltonian corresponding to different charge states, to second order in the $t_{ij}$, by using a Schrieffer-Wolff transformation~\cite{SUPSchrieffer1966}.
We then write the resulting effective Hamiltonian for the sector describing the (1,1,1,1) charge state, focusing on the states with $S^z_{\rm tot} = 0$,
\begin{align} \label{eq:sham}
	\hat H_{\rm eff} = \left( \begin{array}{ccc}
		\hat H_{\rm qubit} & 0 & 0 \\
		0 & \hat H_T & 0 \\
		0 & 0 & \hat H_Q \end{array} \right),
\end{align}
written in the basis $\{\ket{S^\alpha},\ket{S^\beta},\ket{T_0^\alpha},\ket{T_0^\beta},\ket{T_0^\gamma},\ket{Q_0}\}$, see (\ref{eq:sa})--(\ref{eq:q0})~\cite{SUPNote1}.
Since the total Hamiltonian $\hat H_{\rm es} + \hat H_{\rm Z} + \hat H_{\rm t}$ commutes with the total spin projection operator $\hat S^z_{\rm tot}$, none of these six states is coupled to any state with $S^z_{\rm tot}\neq 0$, to all orders in the $t_{ij}$.

The three non-zero sub-blocks in (\ref{eq:sham}) read explicitly
\begin{align}
\hat H_\text{qubit}= {} & {} \left(
    \begin{array}{c c}
       - \frac{1}{4} (J_{12}+J_{24}+4J_{23})  &     \frac{\sqrt 3}{4} (J_{12} - J_{24})           \\
      \frac{\sqrt 3}{4}(J_{12} - J_{24})       &  - \frac{3}{4} (J_{12} + J_{24})  
    \end{array} \right), \\
\hat H_T= {} & {} \left(
    \begin{array}{c c c}
       -\frac{1}{12}(J_{12}+J_{24}+4J_{23})  &    -\frac{1}{4\sqrt 3}(J_{12}-J_{24})   &   \frac{1}{3\sqrt 2}(J_{12}+J_{24}-2J_{23})  \\
           -\frac{1}{4\sqrt 3}(J_{12}-J_{24}) & -\frac{1}{4}(J_{12}+J_{24}) &   -\frac{1}{\sqrt 6}(J_{12}-J_{24})   \\
       \frac{1}{3\sqrt 2}(J_{12}+J_{24}-2J_{23}) &   -\frac{1}{\sqrt 6}(J_{12}-J_{24})  & -\frac{2}{3}(J_{12}+J_{24}+J_{23})
    \end{array} \right), \\
\hat H_Q = {} & {} 0,
\end{align}
where we used the exchange energies
\begin{align}
J_{12} = {} & {} \frac{t_{12}^2}{\Delta + 2\epsilon_\Lambda+\epsilon_{14}-\epsilon_{23}}, \\
J_{24} = {} & {} \frac{t_{24}^2}{\Delta + 2\epsilon_\Lambda-\epsilon_{14}-\epsilon_{23}}, \\
J_{23} = {} & {} \frac{t_{23}^2}{\Delta - 2\epsilon_{23}}.
\end{align}
We see that after subtracting a constant offset this yields the qubit Hamiltonian as presented in Eq.~(4) of the main text~\cite{SUPNote2}, as well as an effective Hamiltonian for the $T_0$-subspace.

For the case where $t_{12} = t_{24} \neq t_{23}$, the Hamiltonian $\hat H_T$ simplifies considerably at the point $Q$ (where it follows that $J_{12} = J_{24}$), and the eigenenergies of $\hat H_T$ are found straightforwardly,
\begin{align}
	E_{1}(Q) = {} & {}   - \frac{1}{2} J-j, \label{eq:e1}\\
	E_{0}(Q) = {} & {}   - \frac{3}{2} J, \\
	E_{T1}(Q) = {} & {} - \frac{3}{4} J - \frac{1}{2} j + \frac{1}{4} \sqrt{9 J^2 - 4 jJ + 4 j^2}, \\
	E_{T2}(Q) = {} & {} - \frac{1}{2} J, \\
	E_{T3}(Q) = {} & {} - \frac{3}{4} J - \frac{1}{2} j - \frac{1}{4} \sqrt{9 J^2 - 4 jJ + 4 j^2}, \label{eq:S40} \\
      E_Q(Q) = {} & {} 0\label{eq:eq},
\end{align}
where $J = t^2_{12}/\Delta = t^2_{24}/\Delta$ and $j=t_{23}^2/\Delta$.
Ignoring corrections to the wave functions of the order $t/\Delta$, the states $\ket{S^\alpha}$, $\ket{S^\beta}$, $\ket{T_2}$, and $\ket{Q_0}$ are eigenstates of $\hat H_\text{eff}$ at point $Q$.
The two remaining $T_0$ eigenstates read
\begin{align}
	\ket{T_1} = {} & {} \cos \frac{\theta}{2} \ket{\Psi_{T\alpha}} + \sin \frac{\theta}{2} \ket{\Psi_{T\gamma}}, \\
	\ket{T_3} = {} & {} \sin \frac{\theta}{2} \ket{\Psi_{T\alpha}} - \cos \frac{\theta}{2} \ket{\Psi_{T\gamma}},
\end{align}
with the angle $\theta$ defined by $\tan \theta = 4\sqrt 2 (J-j)/(7J+2j)$.

The first-order corrections in $t/\Delta$ to these basis states read explicitly
\begin{align}
	\ket{S^\alpha}^{(1)} & = \frac{1}{2}\frac{t}{\Delta} \big[ \ket{S^{(11)}S^{(34)}} - \ket{S^{(13)}S^{(44)}} \big] - \frac{t_{23}}{\Delta} \ket{S^{(14)}S^{(33)}}, \label{eq:sa1}\\
	\ket{S^\beta}^{(1)} & = -\frac{\sqrt 3}{2}\frac{t}{\Delta}\big[ \ket{S^{(11)}S^{(34)}} + \ket{S^{(13)}S^{(44)}} \big] ,\label{eq:sb1}\\
	\ket{T_1}^{(1)} & = -\left(\frac{\cos \frac{\theta}{2}}{2\sqrt{3}} + \frac{2\sin \frac{\theta}{2}}{\sqrt{6}} \right) \frac{t}{\Delta} \big[ \ket{S^{(11)}T_0^{(34)}} - \ket{T_0^{(13)}S^{(44)}} \big]
	-\left( \frac{\cos \frac{\theta}{2}}{\sqrt 3} - \frac{2\sin \frac{\theta}{2}}{\sqrt{6}}\right)
	\frac{t_{23}}{\Delta} \ket{T_0^{(14)}S^{(33)}}, \\
	\ket{T_2}^{(1)} & = -\frac{1}{2} \frac{t}{\Delta} \big[\ket{S^{(11)}T_0^{(34)}} + \ket{T_0^{(13)}S^{(44)}} \big], \\
	\ket{T_3}^{(1)} & =-\left(\frac{\sin \frac{\theta}{2}}{2\sqrt{3}} - \frac{2\cos \frac{\theta}{2}}{\sqrt{6}} \right) \frac{t}{\Delta} \big[ \ket{S^{(11)}T_0^{(34)}} - \ket{T_0^{(13)}S^{(44)}} \big]
	-\left( \frac{\sin \frac{\theta}{2}}{\sqrt 3} + \frac{2\cos \frac{\theta}{2}}{\sqrt{6}}\right)
	\frac{t_{23}}{\Delta} \ket{T_0^{(14)}S^{(33)}}, \\
	\ket{Q_0}^{(1)} & = 0,
\end{align}
where we again used that $t_{12} = t_{24} \equiv t$.
These explicit expressions are needed when calculating the phonon-mediated qubit relaxation rate, see below.

\section{Hyperfine interaction}

\subsection{Effective hyperfine Hamiltonian}

Hyperfine interaction couples the spin of the localized electrons to the nuclear spins of the atoms inside the quantum dot.
Due to the large number of nuclear spins in each dot ($\sim 10^5$--$10^6$), we can model the effect of these nuclear spin baths to first approximation as local effective Zeeman fields acting on the electrons.
We thus describe the hyperfine interaction with the Hamiltonian
\begin{align} \label{eq:shhf0}
	\hat{H}_\text{hf} = \frac{g \mu_B}{2} \sum_{i,\alpha,\beta} \hat{c}^\dagger_{i,\alpha} \mathbf{K}_i \cdot \hat{\boldsymbol{\sigma}}_{\alpha \beta} \hat{c}_{i,\beta},
\end{align}
where $\mathbf{K}_i$ is the local effective nuclear field acting on an electron in dot $i$.
Projecting this Hamiltonian to the (1,1,1,1) and $S^z_\text{tot} = 0$ subspace $\{\ket{S^\alpha},\ket{S^\beta},\ket{T_1},\ket{T_2},\ket{T_3},\ket{Q_0}\}$ yields
\begin{align}\label{eq:shhf1}
	\hat{H}_\text{hf}' = \frac{g \mu_B}{9} \left(
\begin{array}{cccccc}
 0 & 0 & 3 \sqrt{3} \kappa_{14} u & 3 \kappa_{23} & 3 \sqrt{3} \kappa_{14} v & 0 \\
 0 & 0 & \kappa_{23} u-6 \sqrt{2} \kappa_\Lambda  v & 3 \sqrt{3} \kappa_{14} & 6 \sqrt{2} \kappa_\Lambda  u+\kappa_{23} v & 0 \\
 3 \sqrt{3} \kappa_{14} u & \kappa_{23} u-6 \sqrt{2} \kappa_\Lambda  v & 0 & 0 & 0 & \sqrt{2} \kappa_{23} u+6 \kappa_\Lambda  v \\
 3 \kappa_{23} & 3 \sqrt{3} \kappa_{14} & 0 & 0 & 0 & 3 \sqrt{6} \kappa_{14} \\
 3 \sqrt{3} \kappa_{14} v & 6 \sqrt{2} \kappa_\Lambda  u+\kappa_{23} v & 0 & 0 & 0 & \sqrt{2} \kappa_{23} v-6 \kappa_\Lambda  u \\
 0 & 0 & \sqrt{2} \kappa_{23} u+6 \kappa_\Lambda  v & 3 \sqrt{6} \kappa_{14} & \sqrt{2} \kappa_{23} v-6 \kappa_\Lambda  u & 0 \\
\end{array}
\right),
\end{align}
where we introduced the field gradients
\begin{align}
\kappa_{14} = {} & {} g\mu_B \frac{K_{z,1} - K_{z,4}}{2}, \\
\kappa_{23} = {} & {} g\mu_B \frac{K_{z,2} - K_{z,3}}{2}, \\
\kappa_\Lambda = {} & {} g\mu_B \frac{K_{z,1} - K_{z,2} - K_{z,3} + K_{z,4}}{4},
\end{align}
and use the notation
\begin{align}
u = {} & {} \cos \frac{\theta}{2} - \sqrt 2 \sin \frac{\theta}{2}, \\
v = {} & {} \sqrt 2 \cos \frac{\theta}{2} + \sin \frac{\theta}{2}.
\end{align}

We note that:
(i) The projected effective hyperfine Hamiltonian (\ref{eq:shhf1}) does not affect the qubit splitting $E_1 - E_0$ to linear order in the ${\bf K}_i$.
(ii) Since all basis states have $S^z_\text{tot} = 0$, only the $z$-components $K_{z,i}$ appear in the projected Hamiltonian; the perpendicular components $K_{x,i}$ and $K_{y,i}$ couple these states to states with $S^z_\text{tot} \neq 0$, which are split off by the large Zeeman energy $E_{\rm Z}$.
(iii) The field gradients $\kappa_{14}$, $\kappa_{23}$, and $\kappa_\Lambda$ couple only states that differ in total spin $S$ by $\pm 1$, i.e., singlet states are only coupled to triplet states and the quintuplet state only to the triplet states.

\subsection{Higher-order effects}

Although the singlet-only qubit is thus to lowest-order insensitive to the nuclear fields, we can see from the hyperfine Hamiltonian (\ref{eq:shhf1}) that higher-order effects will still lead to a shift in the qubit splitting: to leading order $E_1$ and $E_0$ are both randomly shifted by an energy $\sim K^2/E_J$, where $E_J \sim J,j$ is the typical scale of the exchange energies, which set the splitting between the qubit states and the triplet states, see Eqs.~(\ref{eq:e1})--(\ref{eq:eq}).

A second-order perturbation theory in the nuclear fields yields the energy shifts
\begin{align}
\delta E_1 = {} & {} -\frac{1}{3} \frac{\kappa_{14}^2u^2}{E_{T1}-E_1} - \frac{1}{9} \frac{\kappa_{23}^2}{E_{T2}-E_1} + \frac{1}{3} \frac{\kappa_{14}^2v^2}{E_1-E_{T3}}, \\
\delta E_0 = {} & {} -\frac{1}{81} \frac{(\kappa_{23}u-6\sqrt 2\kappa_\Lambda v)^2}{E_{T1}-E_0} - \frac{1}{3}\frac{\kappa_{14}^2}{E_{T2}-E_0} + \frac{1}{81} \frac{(\kappa_{23} v + 6\sqrt 2 \kappa_\Lambda u)^2}{E_0 - E_{T3}}.
\end{align}
Setting $t_{12} = t_{24} = t$ and $t_{23} = \tfrac{3}{4}t$, as in the main text, this yields
\begin{align}
\delta E_1 - \delta E_0 = \frac{\Delta}{t^2} \left(
\frac{89 \kappa_{14}^2}{72}-\frac{688 \kappa_{23}^2}{2187}-\frac{928 \kappa_{23} \kappa_\Lambda }{729}-\frac{112 \kappa_\Lambda^2}{243} \right),
\end{align}
indeed quadratic in the nuclear fields and of the order of magnitude $\sim K^2 \Delta/t^2$, which is smaller by a factor $\sim K\Delta / t^2$ as compared to the hyperfine-induced level shift in other semiconductor-based spin qubits.
For typical values $K = 1$~mT and $t^2/\Delta = 2~\mu$eV this factor is $0.012$.

\subsection{Hyperfine-induced dephasing of the qubit}

We expect the main effect of these random higher-order corrections to the qubit splitting to be dephasing of the qubit, analogously to how linear shifts $\propto K$ lead to dephasing in single-, double-, and triple-dot spin qubits.
This can be explicitly shown by including the random shift $\delta_{\rm hf} = \delta E_1 - \delta E_0$ into the effective Hamiltonian (5) of the main text,
\begin{align} \label{eq:sheff}
	\hat{H}_\text{eff} = \frac{1}{2} (\hbar \omega_0 +\delta_{\rm hf}) \hat{\sigma}_z + \tilde{A} \cos(\omega_0 \tau) \hat{\sigma}_x,
\end{align}
where we assumed resonant driving, $\epsilon_{14} = A \cos \omega_0\tau$, and defined $\tilde{A}=\sqrt{3}t^2 A/(2 \Delta^2)$.
Using a rotating wave approximation, we write the Hamiltonian in a rotating frame,
\begin{align} \label{eq:sheffrf}
	\hat{H}_\text{rf} = \frac{1}{2} \delta_{\rm hf} \hat{\sigma}_z + \frac{1}{2} \tilde{A} \hat{\sigma}_x.
\end{align}
Assuming a fixed $\delta_{\rm hf}$ and an initial state $\ket{0}$, the resulting Schr\"odinger equation can be solved analytically, yielding the well-known Rabi oscillations of the probability of finding the system in $|1\rangle$ after a time $\tau$,
\begin{align}
	P_1 (\tau,\delta_{\rm hf}) = |\langle 1 | \psi(\tau) \rangle |^2 = \frac{1}{2} \frac{\tilde{A}^2}{\tilde{A}^2 + \delta_{\rm hf}^2}  \left[ 1- \cos\left( \tau \sqrt{\tilde{A}^2 + \delta_{\rm hf}^2}\right) \right],
\end{align}
where we set $\hbar = 1$ for convenience.

We see that the Rabi frequency $\sqrt{\tilde{A}^2 + \delta_{\rm hf}^2}$ depends on the hyperfine-induced shift of the qubit splitting.
The time scale of typical fluctuations in $\delta_{\rm hf}$ is relatively long (up to $\sim 100~\mu$s), which justifies assuming $\delta_{\rm hf}$ to be constant during a single experimental cycle.
Averaging over many cycles thus leads to a suppression of the Rabi oscillations, which can be described as
\begin{align}
\langle P_1(\tau) \rangle =
\int \frac{\text{d}K_{z,1} \text{d}K_{z,2} \text{d}K_{z,3} \text{d}K_{z,4}}{8 \pi^2 \sigma_{K_z}^4} 
\frac{\tilde{A}^2}{\tilde{A}^2 + \delta_{\rm hf}^2}  \left[ 1- \cos\left( \tau \sqrt{\tilde{A}^2 + \delta_{\rm hf}^2}\right) \right]
\exp\left(-\frac{K_{z,1}^2 + K_{z,2}^2 + K_{z,3}^2 + K_{z,4}^2}{2\sigma_{K_z}^2}\right),
\end{align}
where we assumed all $z$-components of the nuclear fields $K_{z,i}$ to be randomly drawn from the same Gaussian distribution centered around zero and having a standard deviation $\sigma_{K_z}$.
A numerical investigation of this integral at large times $\tau \gg \tilde A/\delta^2_{\rm hf}$ suggests that this contribution to qubit dephasing leads to a power-law decay in the Rabi oscillations, $\langle P_1(\tau)\rangle \propto \tau^{-\alpha}$ with $\alpha \approx 1$.

\section{Qubit relaxation due to electron-phonon coupling}

The main source of qubit relaxation, i.e.~dissipative transitions from $\ket{1}$ to $\ket{0}$, is expected to be electron-phonon coupling~\cite{SUPTaylor}.
To estimate the associated relaxation rate, we use a model Hamiltonian for the electron-phonon coupling
\begin{align}
\hat{H}_\text{e-ph} = \sum_{\mathbf{k},p} \lambda_{\mathbf{k},p} \hat{\rho}_{\mathbf{k}} [ \hat{a}_{\mathbf{k},p} + \hat{a}^\dagger_{-\mathbf{k},p} ],
\end{align}
where $\hat{\rho}_{\mathbf{k}}=\int \text{d}\mathbf{r}e^{-i\mathbf{k\cdot r}} \hat{\rho}(\mathbf{r})$ is the Fourier transform of the electronic density operator and $\hat{a}^\dagger_{\mathbf{k},p}$ creates a phonon with momentum $\mathbf{k}$ and polarization $p$. The modulo square of the matrix elements $\lambda_{\mathbf{k},p}$ reads
\begin{align}
\left| \lambda_{\vk,p} \right|^2 =  \frac{\hbar^2 \pi^2 v^2_p}{k \mathcal{V}} \left[ g^{(p)}_\text{pe} \left( A_{\vk,p} \right)^2 + g_\text{def} \sigma^2 k^2 \delta_{p,l} \right],
\end{align}
where $v_p$ is the polarization-dependent sound velocity in the crystal, $\mathcal{V}$ is the normalization volume, and $\sigma$ is the size (radius) of the quantum dots.
We further used the dimensionless coupling constants
\begin{align}
g^{(p)}_\text{pe} \equiv&     \frac{(eh_{14})^2}{2 \pi^2 \hbar \rho_0 v^3_p}, \\
g_\text{def}      \equiv&     \frac{\Xi^2}{2 \pi^2 \hbar \rho_0 v^3_p \sigma^2},
\end{align}
accounting for coupling to piezoelectric phonons and the deformation potential respectively.
Here, $e$ is the elementary electric charge, $h_{14}$ is the piezoelectric constant, $\Xi$ is the deformation potential, and $\rho_0$ is the mass density of the crystal.
The functions $A_{\vk,p}$ are the anisotropy factors, which depend only on the phonon polarization $p=\{l,t_1,t_2\}$ and the direction of the vector $\vk$. For GaAs they read explicitly~\cite{SUPDanon2013}
\begin{align}
(A_{\vk,l})^2   =&    9 \cos^2(\theta) \sin^4(\theta) \sin^2(2\phi), \\
(A_{\vk,t1})^2  =&    \frac{1}{4} \left[ 1+3\cos(2\theta) \right]^2 \sin^2(\theta) \sin^2(2\phi),\\
(A_{\vk,t2})^2  =&    \sin^2(2\theta) \cos^2(2\phi),
\end{align}
where the polar direction $\theta=0$ is taken to be perpendicular to the quadruple-dot plane and the azimuthal angle $\phi$ is the in-plane angle, with $\phi=0$ corresponding to the direction along the 1-2-4 axis [see Fig.~1(a) in the main text].
We assumed that the direction $\{\theta,\phi\} = \{\pi/2,0\}$, i.e.~the 1-2-4 axis, points along the (100) crystallographic direction.

We would now like to find an explicit relaxation rate at the point $Q$ (see Fig.~1 in the main text).
For this purpose we use Fermi's golden rule,
\begin{align}
\Gamma_\text{rel} {} & {} =\frac{2\pi}{\hbar} \sum_{{\bf k},p} \big | \bra{0}\otimes\bra{1_{-{\bf k},p}}  \lambda_{\mathbf{k},p} \hat{\rho}_{\mathbf{k}} \hat{a}^\dagger_{-\mathbf{k},p}  \ket{{\rm vac}} \otimes \ket{1} \big |^2 \delta(E_0 + \hbar\omega_{-{\bf k},p} -E_1) \nonumber \\
{} & {} =\frac{2\pi}{\hbar} \sum_{{\bf k},p} |\lambda_{\mathbf{k},p}|^2 \big | \bra{0} \hat{\rho}_{\mathbf{k}} \ket{1} \big |^2 \delta(E_0 + \hbar\omega_{-{\bf k},p} -E_1),
\end{align}
where the initial state has the qubit in $\ket{1}$ combined with the phonon vacuum, and the final state has the qubit in $\ket{0}$ and one phonon created with wave vector $-{\bf k}$ and polarization $p$.
The energy splitting $E_1 - E_0$ equals the qubit splitting and $\hbar\omega_{-{\bf k},p}$ is the energy of the created phonon.
Using a second-quantization formalism, we write the electronic density operator as
\begin{align}
\hat{\rho}_{\vk} = \sum_{\alpha,i,j} \int \text{d}\vr\, e^{-i\vk \cdot \vr} \psi_i^* (\vr) \psi_j (\vr) \hat{c}^\dagger_{i\alpha} \hat{c}_{j\alpha},
\end{align}
where $\alpha \in \{\ua,\da\}$ labels spin and $i$ and $j$ are summed over the orbital states involved.
We approximate the wave function of an electron localized in dot $i$ by a Gaussian $\psi_i (\vr) = e^{-(\vr-{\bf R}_i)^2/2\sigma^2}/\sqrt{\pi\sigma^2}$, where $\sigma$ sets the spatial extent of the wave function and ${\bf R}_i$ points at the location of the center of the dot.
Neglecting the exponentially suppressed overlap of the wave functions of two electrons localized in different dots, we focus only on terms with $i=j$.
We then use the explicit wave functions (\ref{eq:sa}) and (\ref{eq:sb}) and the corrections (\ref{eq:sa1}) and (\ref{eq:sb1}) at the qubit operation point $Q$, as found above.
This yields to leading order in $t$~\cite{SUPNote3}
\begin{align}
\Gamma_\text{rel} 
{} & {} =\frac{2\pi}{\hbar} \sum_{{\bf k},p} |\lambda_{\mathbf{k},p}|^2
\frac{3}{4} \frac{t^4}{\Delta^4} \sin^2 (k_x d) \exp \left(-\frac{k^2\sigma^2}{2} \right)
\delta(E_0 + \hbar\omega_{-{\bf k},p} -E_1),
\end{align}
where $d$ denotes the distance between two neighboring dots (assumed to be equal for all three ``legs'' of the T), and we assumed that $t_{12} = t_{24} = t$ and $\Delta = U-3U_c$ as before.

For typical qubit splittings (of the order $\sim \mu$eV) the coupling to piezoelectric phonons dominates~\cite{SUPDanon2013}, so we drop the contribution from the deformation potential.
We then convert the sum over ${\bf k}$ to an integral, yielding
\begin{align}
\Gamma_\text{rel} 
{} & {} = \frac{3 \omega_0}{16} \frac{t^4}{\Delta^4} \sum_p g_{\rm pe}^{(p)}
\int d\theta\, d\phi\, \sin\theta \, \sin^2\left( \frac{\omega_0 d \sin\theta \cos \phi}{v_p} \right)
[A_p(\theta,\phi)]^2 \exp \left(- \frac{\omega_0^2 \sigma^2 \sin^2\theta}{2 v^2_p} \right),
\end{align}
where we used that the factors $A_{{\bf k},p}$ only depend on the direction of ${\bf k}$ and introduced the qubit level splitting $\hbar\omega_0 = E_1 - E_0$.
The phonon velocities in GaAs are roughly $v_l = 5.3$~km/s and $v_{t1} = v_{t2} = 3.4$~km/s, which yields
\begin{align}
\frac{\omega_0 d}{v_p}, \frac{\omega_0 \sigma}{v_p} \lesssim 10^{-2},
\end{align}
using that $\hbar\omega_0 \sim \mu$eV and assuming the realistic device parameters $d = 100$~nm and $\sigma = 20$~nm.
The fact that these dimensionless parameters are so small allows us to expand the integrand, which yields to leading order
\begin{align}
\Gamma_\text{rel} 
{} & {} = \frac{3\pi}{105} \omega_0^3 \frac{t^4}{\Delta^4} \left[
4\frac{d^2 g_{\rm pe}^{(t)}}{v_t^2} + 3 \frac{d^2g_{\rm pe}^{(l)}}{v_l^2} \right].
\end{align}
To arrive at the order-of-magnitude estimate given in the main text, we have set $v_t = v_l \equiv v = 4.0$~km/s.

\section{Qubit initialization scheme}

One important ingredient of every quantum computation protocol is the ability to reliably initialize a qubit in one of its basis states.
In analogy to the initialization schemes successfully implemented in double-dot $S$-$T_0$ and triple-dot XO qubits, we propose to use a type of spin-to-charge conversion where the system is pulsed to a point in gate space where there is a unique ground state which has the same spin structure as one of the two qubit states.
After waiting long enough at this initialization point the system will have relaxed to this ground state, and subsequent pulsing back to the qubit operation point yields a qubit uniquely prepared in one of its basis states.

\begin{figure}[t]
      \includegraphics[width=0.97\linewidth]{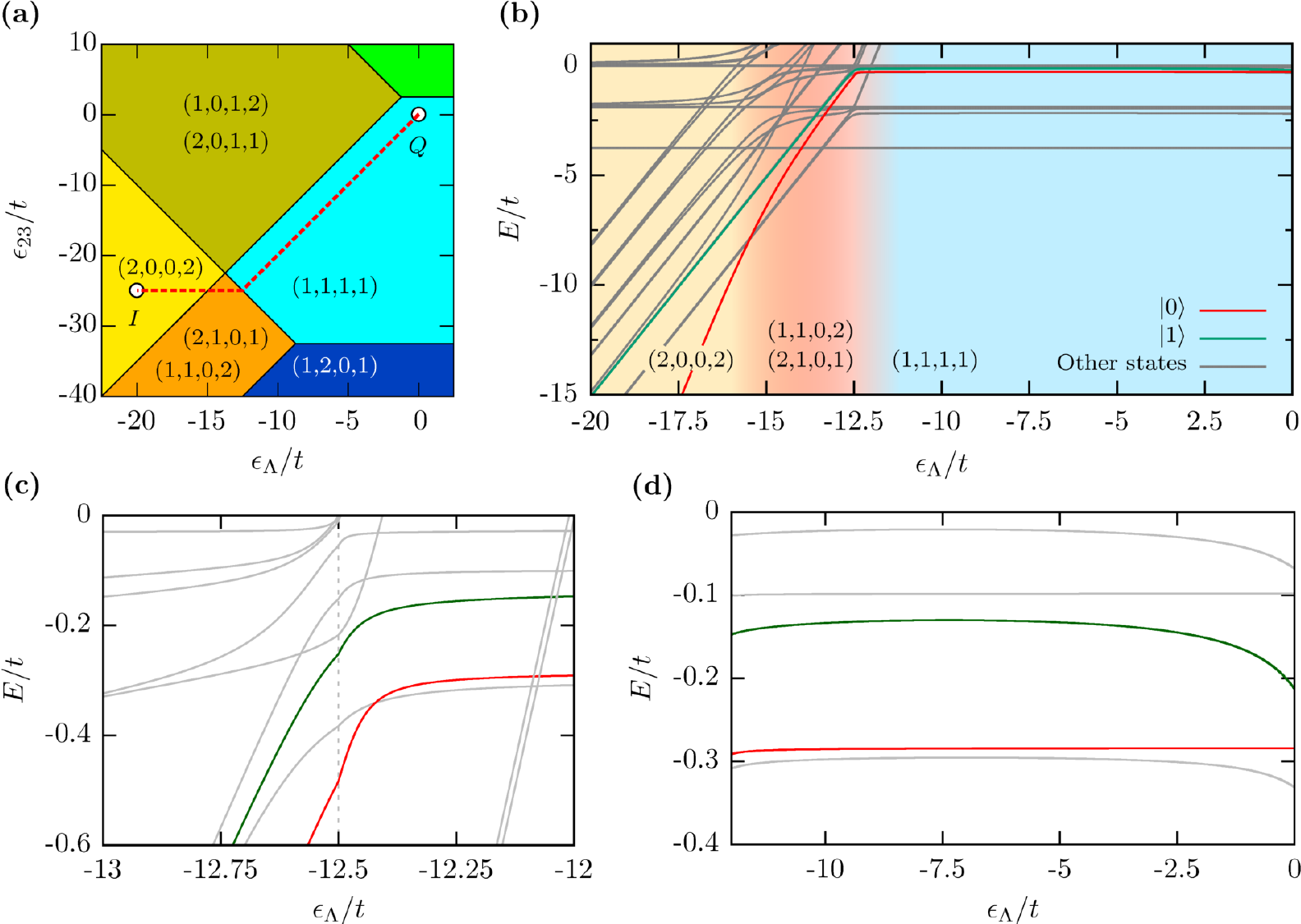}
      \caption{Initialization protocol for the qubit. (a) Charge stability diagram as a function of $\epsilon_\Lambda$ and $\epsilon_{23}$, with $\epsilon_{14}=0$.
            The system is pulsed to the point marked $I$, where it relaxes into its singlet-only (2,0,0,2) ground state.
            Then the red dashed path is followed to bring the system to the qubit operation point $Q$, yielding a qubit initialized in $\ket{0}$.
            (b) Low-energy part of the spectrum along the red dashed line in (a), as a function of $\epsilon_\Lambda$.
            Different regions are colored according to the local ground state charge.
            The red (green) line shows the level adiabatically coupled to $\ket{0}$ ($\ket{1}$).
            (c,d) A zoom-in of the relevant dense parts in (b).\label{fig:s1}}
\end{figure}

In our quadruple-dot singlet-only setup, this can be achieved by pulsing the system to the (2,0,0,2) charge region, e.g.~to the point $I$ in Fig.~\ref{fig:s1}(a).
There the unique ground state consists of two doubly occupied dots where the two ``pairs'' of electrons both have to be in a spin singlet state, due to the large orbital level spacing on the dots.
This ground state can be denoted $|S_{11} S_{44}\rangle$ and when the system is pulsed back to $Q$ along the red dashed line in Fig.~\ref{fig:s1}(a), this state couples adiabatically to the qubit state $\ket{0}$.

In Figs.~\ref{fig:s1}(b--d) we show the lowest part of the full spectrum of the system along the red dashed line in Fig.~\ref{fig:s1}(a), as a function of $\epsilon_\Lambda$ (using the same electrostatic and tunnel coupling parameters as in the main text).
We see that for roughly $\epsilon_\Lambda < 15t$ there is a unique ground state, marked in red, which is indeed the state $\ket{S_{11}S_{44}}$.
Assuming that the system has relaxed to this ground state, subsequent pulsing along the path marked in Fig.~\ref{fig:s1}(a) will make the system follow the state marked in red in Figs.~\ref{fig:s1}(b--d), finally yielding the state $\ket{0}$ when $\epsilon_\Lambda = 0$.

We see that along this path several other levels are crossed, so some caution is required.
The spectrum shown in Fig.~\ref{fig:s1} was calculated using the Hamiltonian $\hat H = \hat H_0 + \hat H_Z + \hat H_t$, as given by Eqs.~(\ref{eq:sham0})-(\ref{eq:shamt}), and this Hamiltonian commutes with the operator $\hat S^z$ as well as the total spin operator $\hat S^2$ of the four-particle spin state.
Since none of the levels that are crossed corresponds to a singlet state, the state marked in red (which is a singlet everywhere) is not coupled to any of them through $\hat H$.
However, other ingredients, such as hyperfine interaction and spin-orbit coupling, do not conserve total spin and can thus couple the singlet to the states with $S^z = 1$.
Roughly speaking, a requirement for successful initialization of the qubit is thus that all level crossings observed in Figs.~\ref{fig:s1}(b--d) are swept through fast enough so that the chance for a hyperfine- or spin-orbit-mediated transition to a different spin state is strongly suppressed.
We estimate the probability for the desired ``diabatic'' transition using the Landau-Zener parameter,
\begin{align}
P_{SS} = \exp \left(-2\pi \frac{W^2}{\hbar v} \right),
\end{align}
where $W$ is the matrix element that couples the two states and $v$ is the sweep speed, i.e.~the rate of change of the energy of the level (in units of energy per second).
We see that this probability approaches unity when $v \gg \pi W^2 / 2\hbar$, and this puts a lower bound on $v$ for successful initialization.
Taking $W = 0.07~\mu$eV, which would be the order of magnitude of a matrix element present due to a set of local random effective magnetic fields of magnitude $\sim 3$~mT, we find as condition $v \gg 0.01~\mu$eV/ns.
Note that a similar constraint holds for the initialization procedure of the triple-dot exchange-only qubit:
Along the path of the initialization pulse, the level crossing with the $S^z = \tfrac{3}{2}$ quadruplet state has to be swept through fast enough to avoid transitions into that state~\cite{SUPMedford}.

Note also that the path we chose for the initialization pulse avoids the charge states (1,0,1,2) and (2,0,1,1).
In these charge states the singly occupied dots are not neighboring each other directly, leading to a reduced singlet-triplet splitting due to interdot exchange effects.
All levels with $S^z = 0$ thus lie much closer to each other, making it harder to sweep through level crossings fast enough and avoid leakage into the triplet states.
A similar problem arises for paths that enter (or come too close to) the charge state (1,2,0,1).

\section{Feasibility of qubit design}

\begin{figure}
	\includegraphics[width=0.35\linewidth]{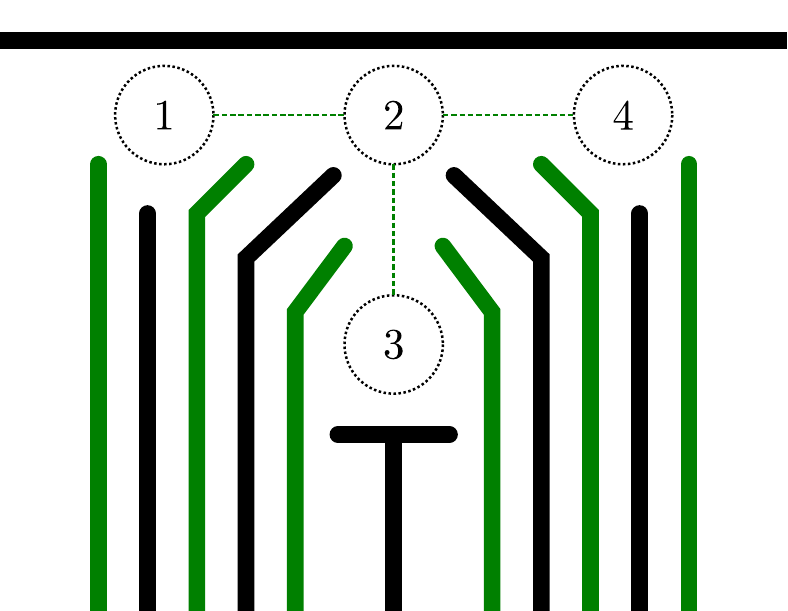}
	\caption{Sketch of a possible gating pattern yielding a quadruple quantum dot in a T setup. Such a design would allow for full independent control of all on-site quantum-dot potentials (black gate electrodes) as well as interdot tunnel coupling energies (green gate electrodes).\label{fig:s2}}
\end{figure}

Current technology allows for the fabrication of relatively complex structures involving four and more quantum dots in gated semiconductor substrates~\cite{SUPThalineau2012,SUPTakakura2014a,SUPDelbecq2014,SUPZajac2016}.
Therefore, although challenging, a device consisting of four dots in a T-like setup can most likely be fabricated without major difficulties.
A possible gating pattern resulting in such a quadruple dot is sketched in Fig.~\ref{fig:s2}, where the thick lines correspond to electrostatic gates and electrons are expected to localize at the positions of the numbered circles.
We see that with this design, all on-site quantum-dot potentials as well as interdot tunnel coupling energies could be tuned independently, controlled by the black and green gate electrodes respectively.

\begin{thebibliography}{50}%
\makeatletter
\providecommand \@ifxundefined [1]{%
 \@ifx{#1\undefined}
}%
\providecommand \@ifnum [1]{%
 \ifnum #1\expandafter \@firstoftwo
 \else \expandafter \@secondoftwo
 \fi
}%
\providecommand \@ifx [1]{%
 \ifx #1\expandafter \@firstoftwo
 \else \expandafter \@secondoftwo
 \fi
}%
\providecommand \natexlab [1]{#1}%
\providecommand \enquote  [1]{``#1''}%
\providecommand \bibnamefont  [1]{#1}%
\providecommand \bibfnamefont [1]{#1}%
\providecommand \citenamefont [1]{#1}%
\providecommand \href@noop [0]{\@secondoftwo}%
\providecommand \href [0]{\begingroup \@sanitize@url \@href}%
\providecommand \@href[1]{\@@startlink{#1}\@@href}%
\providecommand \@@href[1]{\endgroup#1\@@endlink}%
\providecommand \@sanitize@url [0]{\catcode `\\12\catcode `\$12\catcode
  `\&12\catcode `\#12\catcode `\^12\catcode `\_12\catcode `\%12\relax}%
\providecommand \@@startlink[1]{}%
\providecommand \@@endlink[0]{}%
\providecommand \url  [0]{\begingroup\@sanitize@url \@url }%
\providecommand \@url [1]{\endgroup\@href {#1}{\urlprefix }}%
\providecommand \urlprefix  [0]{URL }%
\providecommand \Eprint [0]{\href }%
\providecommand \doibase [0]{http://dx.doi.org/}%
\providecommand \selectlanguage [0]{\@gobble}%
\providecommand \bibinfo  [0]{\@secondoftwo}%
\providecommand \bibfield  [0]{\@secondoftwo}%
\providecommand \translation [1]{[#1]}%
\providecommand \BibitemOpen [0]{}%
\providecommand \bibitemStop [0]{}%
\providecommand \bibitemNoStop [0]{.\EOS\space}%
\providecommand \EOS [0]{\spacefactor3000\relax}%
\providecommand \BibitemShut  [1]{\csname bibitem#1\endcsname}%
\let\auto@bib@innerbib\@empty
\bibitem [{\citenamefont {Hanson}\ \emph {et~al.}(2007)\citenamefont {Hanson},
  \citenamefont {Kouwenhoven}, \citenamefont {Petta}, \citenamefont {Tarucha},\
  and\ \citenamefont {Vandersypen}}]{RevQD}%
  \BibitemOpen
  \bibfield  {author} {\bibinfo {author} {\bibfnamefont {R.}~\bibnamefont
  {Hanson}}, \bibinfo {author} {\bibfnamefont {L.~P.}\ \bibnamefont
  {Kouwenhoven}}, \bibinfo {author} {\bibfnamefont {J.~R.}\ \bibnamefont
  {Petta}}, \bibinfo {author} {\bibfnamefont {S.}~\bibnamefont {Tarucha}}, \
  and\ \bibinfo {author} {\bibfnamefont {L.~M.~K.}\ \bibnamefont
  {Vandersypen}},\ }\href {\doibase 10.1103/RevModPhys.79.1217} {\bibfield
  {journal} {\bibinfo  {journal} {Rev. Mod. Phys.}\ }\textbf {\bibinfo {volume}
  {79}},\ \bibinfo {pages} {1217} (\bibinfo {year} {2007})}\BibitemShut
  {NoStop}%
\bibitem [{\citenamefont {Loss}\ and\ \citenamefont
  {DiVincenzo}(1998)}]{LossDV}%
  \BibitemOpen
  \bibfield  {author} {\bibinfo {author} {\bibfnamefont {D.}~\bibnamefont
  {Loss}}\ and\ \bibinfo {author} {\bibfnamefont {D.~P.}\ \bibnamefont
  {DiVincenzo}},\ }\href {\doibase 10.1103/PhysRevA.57.120} {\bibfield
  {journal} {\bibinfo  {journal} {Phys. Rev. A}\ }\textbf {\bibinfo {volume}
  {57}},\ \bibinfo {pages} {120} (\bibinfo {year} {1998})}\BibitemShut
  {NoStop}%
\bibitem [{\citenamefont {Koppens}\ \emph {et~al.}(2006)\citenamefont
  {Koppens}, \citenamefont {Buizert}, \citenamefont {Tielrooij}, \citenamefont
  {Vink}, \citenamefont {Nowack}, \citenamefont {Meunier}, \citenamefont
  {Kouwenhoven},\ and\ \citenamefont {Vandersypen}}]{Koppens}%
  \BibitemOpen
  \bibfield  {author} {\bibinfo {author} {\bibfnamefont {F.~H.~L.}\
  \bibnamefont {Koppens}}, \bibinfo {author} {\bibfnamefont {C.}~\bibnamefont
  {Buizert}}, \bibinfo {author} {\bibfnamefont {K.~J.}\ \bibnamefont
  {Tielrooij}}, \bibinfo {author} {\bibfnamefont {I.~T.}\ \bibnamefont {Vink}},
  \bibinfo {author} {\bibfnamefont {K.~C.}\ \bibnamefont {Nowack}}, \bibinfo
  {author} {\bibfnamefont {T.}~\bibnamefont {Meunier}}, \bibinfo {author}
  {\bibfnamefont {L.~P.}\ \bibnamefont {Kouwenhoven}}, \ and\ \bibinfo {author}
  {\bibfnamefont {L.~M.~K.}\ \bibnamefont {Vandersypen}},\ }\href {\doibase
  10.1038/nature05065} {\bibfield  {journal} {\bibinfo  {journal} {Nature}\
  }\textbf {\bibinfo {volume} {442}},\ \bibinfo {pages} {766} (\bibinfo {year}
  {2006})}\BibitemShut {NoStop}%
\bibitem [{\citenamefont {Elzerman}\ \emph {et~al.}(2004)\citenamefont
  {Elzerman}, \citenamefont {Hanson}, \citenamefont {Willems~van Beveren},
  \citenamefont {Witkamp}, \citenamefont {Vandersypen},\ and\ \citenamefont
  {Kouwenhoven}}]{Elzerman}%
  \BibitemOpen
  \bibfield  {author} {\bibinfo {author} {\bibfnamefont {J.~M.}\ \bibnamefont
  {Elzerman}}, \bibinfo {author} {\bibfnamefont {R.}~\bibnamefont {Hanson}},
  \bibinfo {author} {\bibfnamefont {L.~H.}\ \bibnamefont {Willems~van
  Beveren}}, \bibinfo {author} {\bibfnamefont {B.}~\bibnamefont {Witkamp}},
  \bibinfo {author} {\bibfnamefont {L.~M.~K.}\ \bibnamefont {Vandersypen}}, \
  and\ \bibinfo {author} {\bibfnamefont {L.~P.}\ \bibnamefont {Kouwenhoven}},\
  }\href {\doibase 10.1038/nature02693} {\bibfield  {journal} {\bibinfo
  {journal} {Nature}\ }\textbf {\bibinfo {volume} {430}},\ \bibinfo {pages}
  {431} (\bibinfo {year} {2004})}\BibitemShut {NoStop}%
\bibitem [{\citenamefont {Erlingsson}\ and\ \citenamefont
  {Nazarov}(2002)}]{PhysRevB.66.155327}%
  \BibitemOpen
  \bibfield  {author} {\bibinfo {author} {\bibfnamefont {S.~I.}\ \bibnamefont
  {Erlingsson}}\ and\ \bibinfo {author} {\bibfnamefont {Y.~V.}\ \bibnamefont
  {Nazarov}},\ }\href {\doibase 10.1103/PhysRevB.66.155327} {\bibfield
  {journal} {\bibinfo  {journal} {Phys. Rev. B}\ }\textbf {\bibinfo {volume}
  {66}},\ \bibinfo {pages} {155327} (\bibinfo {year} {2002})}\BibitemShut
  {NoStop}%
\bibitem [{\citenamefont {Merkulov}\ \emph {et~al.}(2002)\citenamefont
  {Merkulov}, \citenamefont {Efros},\ and\ \citenamefont {Rosen}}]{Merkulov}%
  \BibitemOpen
  \bibfield  {author} {\bibinfo {author} {\bibfnamefont {I.~A.}\ \bibnamefont
  {Merkulov}}, \bibinfo {author} {\bibfnamefont {A.~L.}\ \bibnamefont {Efros}},
  \ and\ \bibinfo {author} {\bibfnamefont {M.}~\bibnamefont {Rosen}},\ }\href
  {\doibase 10.1103/PhysRevB.65.205309} {\bibfield  {journal} {\bibinfo
  {journal} {Phys. Rev. B}\ }\textbf {\bibinfo {volume} {65}},\ \bibinfo
  {pages} {205309} (\bibinfo {year} {2002})}\BibitemShut {NoStop}%
\bibitem [{\citenamefont {Khaetskii}\ \emph {et~al.}(2002)\citenamefont
  {Khaetskii}, \citenamefont {Loss},\ and\ \citenamefont {Glazman}}]{klg}%
  \BibitemOpen
  \bibfield  {author} {\bibinfo {author} {\bibfnamefont {A.~V.}\ \bibnamefont
  {Khaetskii}}, \bibinfo {author} {\bibfnamefont {D.}~\bibnamefont {Loss}}, \
  and\ \bibinfo {author} {\bibfnamefont {L.}~\bibnamefont {Glazman}},\ }\href
  {\doibase 10.1103/PhysRevLett.88.186802} {\bibfield  {journal} {\bibinfo
  {journal} {Phys. Rev. Lett.}\ }\textbf {\bibinfo {volume} {88}},\ \bibinfo
  {pages} {186802} (\bibinfo {year} {2002})}\BibitemShut {NoStop}%
\bibitem [{\citenamefont {Golovach}\ \emph {et~al.}(2006)\citenamefont
  {Golovach}, \citenamefont {Borhani},\ and\ \citenamefont {Loss}}]{Golovach}%
  \BibitemOpen
  \bibfield  {author} {\bibinfo {author} {\bibfnamefont {V.~N.}\ \bibnamefont
  {Golovach}}, \bibinfo {author} {\bibfnamefont {M.}~\bibnamefont {Borhani}}, \
  and\ \bibinfo {author} {\bibfnamefont {D.}~\bibnamefont {Loss}},\ }\href
  {\doibase 10.1103/PhysRevB.74.165319} {\bibfield  {journal} {\bibinfo
  {journal} {Phys. Rev. B}\ }\textbf {\bibinfo {volume} {74}},\ \bibinfo
  {pages} {165319} (\bibinfo {year} {2006})}\BibitemShut {NoStop}%
\bibitem [{\citenamefont {Flindt}\ \emph {et~al.}(2006)\citenamefont {Flindt},
  \citenamefont {S\o{}rensen},\ and\ \citenamefont {Flensberg}}]{flindt:prl}%
  \BibitemOpen
  \bibfield  {author} {\bibinfo {author} {\bibfnamefont {C.}~\bibnamefont
  {Flindt}}, \bibinfo {author} {\bibfnamefont {A.~S.}\ \bibnamefont
  {S\o{}rensen}}, \ and\ \bibinfo {author} {\bibfnamefont {K.}~\bibnamefont
  {Flensberg}},\ }\href {\doibase 10.1103/PhysRevLett.97.240501} {\bibfield
  {journal} {\bibinfo  {journal} {Phys. Rev. Lett.}\ }\textbf {\bibinfo
  {volume} {97}},\ \bibinfo {pages} {240501} (\bibinfo {year}
  {2006})}\BibitemShut {NoStop}%
\bibitem [{\citenamefont {Nadj-Perge}\ \emph {et~al.}(2010)\citenamefont
  {Nadj-Perge}, \citenamefont {Frolov}, \citenamefont {Bakkers},\ and\
  \citenamefont {Kouwenhoven}}]{Nadj}%
  \BibitemOpen
  \bibfield  {author} {\bibinfo {author} {\bibfnamefont {S.}~\bibnamefont
  {Nadj-Perge}}, \bibinfo {author} {\bibfnamefont {S.~M.}\ \bibnamefont
  {Frolov}}, \bibinfo {author} {\bibfnamefont {E.~P. A.~M.}\ \bibnamefont
  {Bakkers}}, \ and\ \bibinfo {author} {\bibfnamefont {L.~P.}\ \bibnamefont
  {Kouwenhoven}},\ }\href {\doibase 10.1038/nature09682} {\bibfield  {journal}
  {\bibinfo  {journal} {Nature}\ }\textbf {\bibinfo {volume} {468}},\ \bibinfo
  {pages} {1084} (\bibinfo {year} {2010})}\BibitemShut {NoStop}%
\bibitem [{\citenamefont {Khaetskii}\ and\ \citenamefont
  {Nazarov}(2001)}]{Khaetskii2001}%
  \BibitemOpen
  \bibfield  {author} {\bibinfo {author} {\bibfnamefont {A.~V.}\ \bibnamefont
  {Khaetskii}}\ and\ \bibinfo {author} {\bibfnamefont {Y.~V.}\ \bibnamefont
  {Nazarov}},\ }\href {\doibase 10.1103/PhysRevB.64.125316} {\bibfield
  {journal} {\bibinfo  {journal} {Phys. Rev. B}\ }\textbf {\bibinfo {volume}
  {64}},\ \bibinfo {pages} {125316} (\bibinfo {year} {2001})}\BibitemShut
  {NoStop}%
\bibitem [{\citenamefont {Danon}\ and\ \citenamefont
  {Nazarov}(2009)}]{PhysRevB.80.041301}%
  \BibitemOpen
  \bibfield  {author} {\bibinfo {author} {\bibfnamefont {J.}~\bibnamefont
  {Danon}}\ and\ \bibinfo {author} {\bibfnamefont {Y.~V.}\ \bibnamefont
  {Nazarov}},\ }\href {\doibase 10.1103/PhysRevB.80.041301} {\bibfield
  {journal} {\bibinfo  {journal} {Phys. Rev. B}\ }\textbf {\bibinfo {volume}
  {80}},\ \bibinfo {pages} {041301} (\bibinfo {year} {2009})}\BibitemShut
  {NoStop}%
\bibitem [{\citenamefont {DiVincenzo}\ \emph {et~al.}(2000)\citenamefont
  {DiVincenzo}, \citenamefont {Bacon}, \citenamefont {Kempe}, \citenamefont
  {Burkard},\ and\ \citenamefont {Whaley}}]{DiVincenzo2000}%
  \BibitemOpen
  \bibfield  {author} {\bibinfo {author} {\bibfnamefont {D.~P.}\ \bibnamefont
  {DiVincenzo}}, \bibinfo {author} {\bibfnamefont {D.}~\bibnamefont {Bacon}},
  \bibinfo {author} {\bibfnamefont {J.}~\bibnamefont {Kempe}}, \bibinfo
  {author} {\bibfnamefont {G.}~\bibnamefont {Burkard}}, \ and\ \bibinfo
  {author} {\bibfnamefont {K.~B.}\ \bibnamefont {Whaley}},\ }\href {\doibase
  10.1038/35042541} {\bibfield  {journal} {\bibinfo  {journal} {Nature}\
  }\textbf {\bibinfo {volume} {408}},\ \bibinfo {pages} {339} (\bibinfo {year}
  {2000})}\BibitemShut {NoStop}%
\bibitem [{\citenamefont {Petta}\ \emph {et~al.}(2005)\citenamefont {Petta},
  \citenamefont {Johnson}, \citenamefont {Taylor}, \citenamefont {Laird},
  \citenamefont {Yacoby}, \citenamefont {Lukin}, \citenamefont {Marcus},
  \citenamefont {Hanson},\ and\ \citenamefont {Gossard}}]{Petta}%
  \BibitemOpen
  \bibfield  {author} {\bibinfo {author} {\bibfnamefont {J.~R.}\ \bibnamefont
  {Petta}}, \bibinfo {author} {\bibfnamefont {A.~C.}\ \bibnamefont {Johnson}},
  \bibinfo {author} {\bibfnamefont {J.~M.}\ \bibnamefont {Taylor}}, \bibinfo
  {author} {\bibfnamefont {E.~A.}\ \bibnamefont {Laird}}, \bibinfo {author}
  {\bibfnamefont {A.}~\bibnamefont {Yacoby}}, \bibinfo {author} {\bibfnamefont
  {M.~D.}\ \bibnamefont {Lukin}}, \bibinfo {author} {\bibfnamefont {C.~M.}\
  \bibnamefont {Marcus}}, \bibinfo {author} {\bibfnamefont {M.~P.}\
  \bibnamefont {Hanson}}, \ and\ \bibinfo {author} {\bibfnamefont {A.~C.}\
  \bibnamefont {Gossard}},\ }\href {\doibase 10.1126/science.1116955}
  {\bibfield  {journal} {\bibinfo  {journal} {Science}\ }\textbf {\bibinfo
  {volume} {309}},\ \bibinfo {pages} {2180} (\bibinfo {year}
  {2005})}\BibitemShut {NoStop}%
\bibitem [{\citenamefont {Taylor}\ \emph {et~al.}(2007)\citenamefont {Taylor},
  \citenamefont {Petta}, \citenamefont {Johnson}, \citenamefont {Yacoby},
  \citenamefont {Marcus},\ and\ \citenamefont {Lukin}}]{Taylor2007}%
  \BibitemOpen
  \bibfield  {author} {\bibinfo {author} {\bibfnamefont {J.~M.}\ \bibnamefont
  {Taylor}}, \bibinfo {author} {\bibfnamefont {J.~R.}\ \bibnamefont {Petta}},
  \bibinfo {author} {\bibfnamefont {A.~C.}\ \bibnamefont {Johnson}}, \bibinfo
  {author} {\bibfnamefont {A.}~\bibnamefont {Yacoby}}, \bibinfo {author}
  {\bibfnamefont {C.~M.}\ \bibnamefont {Marcus}}, \ and\ \bibinfo {author}
  {\bibfnamefont {M.~D.}\ \bibnamefont {Lukin}},\ }\href {\doibase
  10.1103/PhysRevB.76.035315} {\bibfield  {journal} {\bibinfo  {journal} {Phys.
  Rev. B}\ }\textbf {\bibinfo {volume} {76}},\ \bibinfo {pages} {035315}
  (\bibinfo {year} {2007})}\BibitemShut {NoStop}%
\bibitem [{\citenamefont {Medford}\ \emph
  {et~al.}(2013{\natexlab{a}})\citenamefont {Medford}, \citenamefont {Beil},
  \citenamefont {Taylor}, \citenamefont {Bartlett}, \citenamefont {Doherty},
  \citenamefont {Rashba}, \citenamefont {DiVincenzo}, \citenamefont {Lu},
  \citenamefont {Gossard},\ and\ \citenamefont {Marcus}}]{Medford}%
  \BibitemOpen
  \bibfield  {author} {\bibinfo {author} {\bibfnamefont {J.}~\bibnamefont
  {Medford}}, \bibinfo {author} {\bibfnamefont {J.}~\bibnamefont {Beil}},
  \bibinfo {author} {\bibfnamefont {J.~M.}\ \bibnamefont {Taylor}}, \bibinfo
  {author} {\bibfnamefont {S.~D.}\ \bibnamefont {Bartlett}}, \bibinfo {author}
  {\bibfnamefont {A.~C.}\ \bibnamefont {Doherty}}, \bibinfo {author}
  {\bibfnamefont {E.~I.}\ \bibnamefont {Rashba}}, \bibinfo {author}
  {\bibfnamefont {D.~P.}\ \bibnamefont {DiVincenzo}}, \bibinfo {author}
  {\bibfnamefont {H.}~\bibnamefont {Lu}}, \bibinfo {author} {\bibfnamefont
  {A.~C.}\ \bibnamefont {Gossard}}, \ and\ \bibinfo {author} {\bibfnamefont
  {C.~M.}\ \bibnamefont {Marcus}},\ }\href
  {http://dx.doi.org/10.1038/nnano.2013.168} {\bibfield  {journal} {\bibinfo
  {journal} {Nat. Nano.}\ }\textbf {\bibinfo {volume} {8}},\ \bibinfo {pages}
  {654} (\bibinfo {year} {2013}{\natexlab{a}})}\BibitemShut {NoStop}%
\bibitem [{\citenamefont {Laird}\ \emph {et~al.}(2010)\citenamefont {Laird},
  \citenamefont {Taylor}, \citenamefont {DiVincenzo}, \citenamefont {Marcus},
  \citenamefont {Hanson},\ and\ \citenamefont {Gossard}}]{PhysRevB.82.075403}%
  \BibitemOpen
  \bibfield  {author} {\bibinfo {author} {\bibfnamefont {E.~A.}\ \bibnamefont
  {Laird}}, \bibinfo {author} {\bibfnamefont {J.~M.}\ \bibnamefont {Taylor}},
  \bibinfo {author} {\bibfnamefont {D.~P.}\ \bibnamefont {DiVincenzo}},
  \bibinfo {author} {\bibfnamefont {C.~M.}\ \bibnamefont {Marcus}}, \bibinfo
  {author} {\bibfnamefont {M.~P.}\ \bibnamefont {Hanson}}, \ and\ \bibinfo
  {author} {\bibfnamefont {A.~C.}\ \bibnamefont {Gossard}},\ }\href {\doibase
  10.1103/PhysRevB.82.075403} {\bibfield  {journal} {\bibinfo  {journal} {Phys.
  Rev. B}\ }\textbf {\bibinfo {volume} {82}},\ \bibinfo {pages} {075403}
  (\bibinfo {year} {2010})}\BibitemShut {NoStop}%
\bibitem [{\citenamefont {Russ}\ and\ \citenamefont {Burkard}()}]{Russ}%
  \BibitemOpen
  \bibfield  {author} {\bibinfo {author} {\bibfnamefont {M.}~\bibnamefont
  {Russ}}\ and\ \bibinfo {author} {\bibfnamefont {G.}~\bibnamefont {Burkard}},\
  }\href {http://arxiv.org/abs/1611.09106v1} {\bibinfo  {journal}
  {arXiv:1611.09106v1}}\BibitemShut {NoStop}%
\bibitem [{\citenamefont {Martins}\ \emph {et~al.}(2016)\citenamefont
  {Martins}, \citenamefont {Malinowski}, \citenamefont {Nissen}, \citenamefont
  {Barnes}, \citenamefont {Fallahi}, \citenamefont {Gardner}, \citenamefont
  {Manfra}, \citenamefont {Marcus},\ and\ \citenamefont
  {Kuemmeth}}]{Martins2016}%
  \BibitemOpen
\bibfield  {journal} {  }\bibfield  {author} {\bibinfo {author} {\bibfnamefont
  {F.}~\bibnamefont {Martins}}, \bibinfo {author} {\bibfnamefont {F.~K.}\
  \bibnamefont {Malinowski}}, \bibinfo {author} {\bibfnamefont {P.~D.}\
  \bibnamefont {Nissen}}, \bibinfo {author} {\bibfnamefont {E.}~\bibnamefont
  {Barnes}}, \bibinfo {author} {\bibfnamefont {S.}~\bibnamefont {Fallahi}},
  \bibinfo {author} {\bibfnamefont {G.~C.}\ \bibnamefont {Gardner}}, \bibinfo
  {author} {\bibfnamefont {M.~J.}\ \bibnamefont {Manfra}}, \bibinfo {author}
  {\bibfnamefont {C.~M.}\ \bibnamefont {Marcus}}, \ and\ \bibinfo {author}
  {\bibfnamefont {F.}~\bibnamefont {Kuemmeth}},\ }\href {\doibase
  10.1103/PhysRevLett.116.116801} {\bibfield  {journal} {\bibinfo  {journal}
  {Phys. Rev. Lett.}\ }\textbf {\bibinfo {volume} {116}},\ \bibinfo {pages}
  {116801} (\bibinfo {year} {2016})}\BibitemShut {NoStop}%
\bibitem [{\citenamefont {Reed}\ \emph {et~al.}(2016)\citenamefont {Reed},
  \citenamefont {Maune}, \citenamefont {Andrews}, \citenamefont {Borselli},
  \citenamefont {Eng}, \citenamefont {Jura}, \citenamefont {Kiselev},
  \citenamefont {Ladd}, \citenamefont {Merkel}, \citenamefont {Milosavljevic},
  \citenamefont {Pritchett}, \citenamefont {Rakher}, \citenamefont {Ross},
  \citenamefont {Schmitz}, \citenamefont {Smith}, \citenamefont {Wright},
  \citenamefont {Gyure},\ and\ \citenamefont {Hunter}}]{Reed2016}%
  \BibitemOpen
  \bibfield  {author} {\bibinfo {author} {\bibfnamefont {M.~D.}\ \bibnamefont
  {Reed}}, \bibinfo {author} {\bibfnamefont {B.~M.}\ \bibnamefont {Maune}},
  \bibinfo {author} {\bibfnamefont {R.~W.}\ \bibnamefont {Andrews}}, \bibinfo
  {author} {\bibfnamefont {M.~G.}\ \bibnamefont {Borselli}}, \bibinfo {author}
  {\bibfnamefont {K.}~\bibnamefont {Eng}}, \bibinfo {author} {\bibfnamefont
  {M.~P.}\ \bibnamefont {Jura}}, \bibinfo {author} {\bibfnamefont {A.~A.}\
  \bibnamefont {Kiselev}}, \bibinfo {author} {\bibfnamefont {T.~D.}\
  \bibnamefont {Ladd}}, \bibinfo {author} {\bibfnamefont {S.~T.}\ \bibnamefont
  {Merkel}}, \bibinfo {author} {\bibfnamefont {I.}~\bibnamefont
  {Milosavljevic}}, \bibinfo {author} {\bibfnamefont {E.~J.}\ \bibnamefont
  {Pritchett}}, \bibinfo {author} {\bibfnamefont {M.~T.}\ \bibnamefont
  {Rakher}}, \bibinfo {author} {\bibfnamefont {R.~S.}\ \bibnamefont {Ross}},
  \bibinfo {author} {\bibfnamefont {A.~E.}\ \bibnamefont {Schmitz}}, \bibinfo
  {author} {\bibfnamefont {A.}~\bibnamefont {Smith}}, \bibinfo {author}
  {\bibfnamefont {J.~A.}\ \bibnamefont {Wright}}, \bibinfo {author}
  {\bibfnamefont {M.~F.}\ \bibnamefont {Gyure}}, \ and\ \bibinfo {author}
  {\bibfnamefont {A.~T.}\ \bibnamefont {Hunter}},\ }\href {\doibase
  10.1103/PhysRevLett.116.110402} {\bibfield  {journal} {\bibinfo  {journal}
  {Phys. Rev. Lett.}\ }\textbf {\bibinfo {volume} {116}},\ \bibinfo {pages}
  {110402} (\bibinfo {year} {2016})}\BibitemShut {NoStop}%
\bibitem [{\citenamefont {Shim}\ and\ \citenamefont {Tahan}(2016)}]{Shim2016}%
  \BibitemOpen
  \bibfield  {author} {\bibinfo {author} {\bibfnamefont {Y.-P.}\ \bibnamefont
  {Shim}}\ and\ \bibinfo {author} {\bibfnamefont {C.}~\bibnamefont {Tahan}},\
  }\href {\doibase 10.1103/PhysRevB.93.121410} {\bibfield  {journal} {\bibinfo
  {journal} {Phys. Rev. B}\ }\textbf {\bibinfo {volume} {93}},\ \bibinfo
  {pages} {121410} (\bibinfo {year} {2016})}\BibitemShut {NoStop}%
\bibitem [{\citenamefont {Coish}\ and\ \citenamefont {Loss}(2005)}]{Coish2005}%
  \BibitemOpen
  \bibfield  {author} {\bibinfo {author} {\bibfnamefont {W.~A.}\ \bibnamefont
  {Coish}}\ and\ \bibinfo {author} {\bibfnamefont {D.}~\bibnamefont {Loss}},\
  }\href {\doibase 10.1103/PhysRevB.72.125337} {\bibfield  {journal} {\bibinfo
  {journal} {Phys. Rev. B}\ }\textbf {\bibinfo {volume} {72}},\ \bibinfo
  {pages} {125337} (\bibinfo {year} {2005})}\BibitemShut {NoStop}%
\bibitem [{\citenamefont {Medford}\ \emph
  {et~al.}(2013{\natexlab{b}})\citenamefont {Medford}, \citenamefont {Beil},
  \citenamefont {Taylor}, \citenamefont {Rashba}, \citenamefont {Lu},
  \citenamefont {Gossard},\ and\ \citenamefont {Marcus}}]{Medford2}%
  \BibitemOpen
  \bibfield  {author} {\bibinfo {author} {\bibfnamefont {J.}~\bibnamefont
  {Medford}}, \bibinfo {author} {\bibfnamefont {J.}~\bibnamefont {Beil}},
  \bibinfo {author} {\bibfnamefont {J.~M.}\ \bibnamefont {Taylor}}, \bibinfo
  {author} {\bibfnamefont {E.~I.}\ \bibnamefont {Rashba}}, \bibinfo {author}
  {\bibfnamefont {H.}~\bibnamefont {Lu}}, \bibinfo {author} {\bibfnamefont
  {A.~C.}\ \bibnamefont {Gossard}}, \ and\ \bibinfo {author} {\bibfnamefont
  {C.~M.}\ \bibnamefont {Marcus}},\ }\href {\doibase
  10.1103/PhysRevLett.111.050501} {\bibfield  {journal} {\bibinfo  {journal}
  {Phys. Rev. Lett.}\ }\textbf {\bibinfo {volume} {111}},\ \bibinfo {pages}
  {050501} (\bibinfo {year} {2013}{\natexlab{b}})}\BibitemShut {NoStop}%
\bibitem [{\citenamefont {Mehl}\ and\ \citenamefont
  {DiVincenzo}(2013)}]{Mehl2013}%
  \BibitemOpen
  \bibfield  {author} {\bibinfo {author} {\bibfnamefont {S.}~\bibnamefont
  {Mehl}}\ and\ \bibinfo {author} {\bibfnamefont {D.~P.}\ \bibnamefont
  {DiVincenzo}},\ }\href {\doibase 10.1103/PhysRevB.87.195309} {\bibfield
  {journal} {\bibinfo  {journal} {Phys. Rev. B}\ }\textbf {\bibinfo {volume}
  {87}},\ \bibinfo {pages} {195309} (\bibinfo {year} {2013})}\BibitemShut
  {NoStop}%
\bibitem [{\citenamefont {Hung}\ \emph {et~al.}(2014)\citenamefont {Hung},
  \citenamefont {Fei}, \citenamefont {Friesen},\ and\ \citenamefont
  {Hu}}]{Hung2014}%
  \BibitemOpen
  \bibfield  {author} {\bibinfo {author} {\bibfnamefont {J.-T.}\ \bibnamefont
  {Hung}}, \bibinfo {author} {\bibfnamefont {J.}~\bibnamefont {Fei}}, \bibinfo
  {author} {\bibfnamefont {M.}~\bibnamefont {Friesen}}, \ and\ \bibinfo
  {author} {\bibfnamefont {X.}~\bibnamefont {Hu}},\ }\href {\doibase
  10.1103/PhysRevB.90.045308} {\bibfield  {journal} {\bibinfo  {journal} {Phys
  Rev. B}\ }\textbf {\bibinfo {volume} {90}},\ \bibinfo {pages} {045308}
  (\bibinfo {year} {2014})}\BibitemShut {NoStop}%
\bibitem [{\citenamefont {Rudner}\ and\ \citenamefont
  {Levitov}(2007)}]{rudnercooling}%
  \BibitemOpen
  \bibfield  {author} {\bibinfo {author} {\bibfnamefont {M.~S.}\ \bibnamefont
  {Rudner}}\ and\ \bibinfo {author} {\bibfnamefont {L.~S.}\ \bibnamefont
  {Levitov}},\ }\href {\doibase 10.1103/PhysRevLett.99.036602} {\bibfield
  {journal} {\bibinfo  {journal} {Phys. Rev. Lett.}\ }\textbf {\bibinfo
  {volume} {99}},\ \bibinfo {pages} {036602} (\bibinfo {year}
  {2007})}\BibitemShut {NoStop}%
\bibitem [{\citenamefont {Vink}\ \emph {et~al.}(2009)\citenamefont {Vink},
  \citenamefont {Nowack}, \citenamefont {Koppens}, \citenamefont {Danon},
  \citenamefont {Nazarov},\ and\ \citenamefont {Vandersypen}}]{Vink}%
  \BibitemOpen
  \bibfield  {author} {\bibinfo {author} {\bibfnamefont {I.~T.}\ \bibnamefont
  {Vink}}, \bibinfo {author} {\bibfnamefont {K.~C.}\ \bibnamefont {Nowack}},
  \bibinfo {author} {\bibfnamefont {F.~H.~L.}\ \bibnamefont {Koppens}},
  \bibinfo {author} {\bibfnamefont {J.}~\bibnamefont {Danon}}, \bibinfo
  {author} {\bibfnamefont {Y.~V.}\ \bibnamefont {Nazarov}}, \ and\ \bibinfo
  {author} {\bibfnamefont {L.~M.~K.}\ \bibnamefont {Vandersypen}},\ }\href
  {\doibase 10.1038/nphys1366} {\bibfield  {journal} {\bibinfo  {journal} {Nat
  Phys}\ }\textbf {\bibinfo {volume} {5}},\ \bibinfo {pages} {764} (\bibinfo
  {year} {2009})}\BibitemShut {NoStop}%
\bibitem [{\citenamefont {Rudner}\ \emph {et~al.}(2011)\citenamefont {Rudner},
  \citenamefont {Vandersypen}, \citenamefont {Vuleti{\'{c}}},\ and\
  \citenamefont {Levitov}}]{Rudner2011}%
  \BibitemOpen
  \bibfield  {author} {\bibinfo {author} {\bibfnamefont {M.~S.}\ \bibnamefont
  {Rudner}}, \bibinfo {author} {\bibfnamefont {L.~M.~K.}\ \bibnamefont
  {Vandersypen}}, \bibinfo {author} {\bibfnamefont {V.}~\bibnamefont
  {Vuleti{\'{c}}}}, \ and\ \bibinfo {author} {\bibfnamefont {L.~S.}\
  \bibnamefont {Levitov}},\ }\href {\doibase 10.1103/PhysRevLett.107.206806}
  {\bibfield  {journal} {\bibinfo  {journal} {Phys. Rev. Lett.}\ }\textbf
  {\bibinfo {volume} {107}},\ \bibinfo {pages} {206806} (\bibinfo {year}
  {2011})}\BibitemShut {NoStop}%
\bibitem [{\citenamefont {Frolov}\ \emph {et~al.}(2012)\citenamefont {Frolov},
  \citenamefont {Danon}, \citenamefont {Nadj-Perge}, \citenamefont {Zuo},
  \citenamefont {van Tilburg}, \citenamefont {Pribiag}, \citenamefont {van~den
  Berg}, \citenamefont {Bakkers},\ and\ \citenamefont {Kouwenhoven}}]{Frolov}%
  \BibitemOpen
  \bibfield  {author} {\bibinfo {author} {\bibfnamefont {S.~M.}\ \bibnamefont
  {Frolov}}, \bibinfo {author} {\bibfnamefont {J.}~\bibnamefont {Danon}},
  \bibinfo {author} {\bibfnamefont {S.}~\bibnamefont {Nadj-Perge}}, \bibinfo
  {author} {\bibfnamefont {K.}~\bibnamefont {Zuo}}, \bibinfo {author}
  {\bibfnamefont {J.~W.~W.}\ \bibnamefont {van Tilburg}}, \bibinfo {author}
  {\bibfnamefont {V.~S.}\ \bibnamefont {Pribiag}}, \bibinfo {author}
  {\bibfnamefont {J.~W.~G.}\ \bibnamefont {van~den Berg}}, \bibinfo {author}
  {\bibfnamefont {E.~P. A.~M.}\ \bibnamefont {Bakkers}}, \ and\ \bibinfo
  {author} {\bibfnamefont {L.~P.}\ \bibnamefont {Kouwenhoven}},\ }\href
  {\doibase 10.1103/PhysRevLett.109.236805} {\bibfield  {journal} {\bibinfo
  {journal} {Phys. Rev. Lett.}\ }\textbf {\bibinfo {volume} {109}},\ \bibinfo
  {pages} {236805} (\bibinfo {year} {2012})}\BibitemShut {NoStop}%
\bibitem [{\citenamefont {Bluhm}\ \emph {et~al.}(2010)\citenamefont {Bluhm},
  \citenamefont {Foletti}, \citenamefont {Mahalu}, \citenamefont {Umansky},\
  and\ \citenamefont {Yacoby}}]{Bluhm}%
  \BibitemOpen
  \bibfield  {author} {\bibinfo {author} {\bibfnamefont {H.}~\bibnamefont
  {Bluhm}}, \bibinfo {author} {\bibfnamefont {S.}~\bibnamefont {Foletti}},
  \bibinfo {author} {\bibfnamefont {D.}~\bibnamefont {Mahalu}}, \bibinfo
  {author} {\bibfnamefont {V.}~\bibnamefont {Umansky}}, \ and\ \bibinfo
  {author} {\bibfnamefont {A.}~\bibnamefont {Yacoby}},\ }\href {\doibase
  10.1103/PhysRevLett.105.216803} {\bibfield  {journal} {\bibinfo  {journal}
  {Phys. Rev. Lett.}\ }\textbf {\bibinfo {volume} {105}},\ \bibinfo {pages}
  {216803} (\bibinfo {year} {2010})}\BibitemShut {NoStop}%
\bibitem [{\citenamefont {Bluhm}\ \emph {et~al.}(2011)\citenamefont {Bluhm},
  \citenamefont {Foletti}, \citenamefont {Neder}, \citenamefont {Rudner},
  \citenamefont {Mahalu}, \citenamefont {Umansky},\ and\ \citenamefont
  {Yacoby}}]{Bluhm2010a}%
  \BibitemOpen
  \bibfield  {author} {\bibinfo {author} {\bibfnamefont {H.}~\bibnamefont
  {Bluhm}}, \bibinfo {author} {\bibfnamefont {S.}~\bibnamefont {Foletti}},
  \bibinfo {author} {\bibfnamefont {I.}~\bibnamefont {Neder}}, \bibinfo
  {author} {\bibfnamefont {M.}~\bibnamefont {Rudner}}, \bibinfo {author}
  {\bibfnamefont {D.}~\bibnamefont {Mahalu}}, \bibinfo {author} {\bibfnamefont
  {V.}~\bibnamefont {Umansky}}, \ and\ \bibinfo {author} {\bibfnamefont
  {A.}~\bibnamefont {Yacoby}},\ }\href {\doibase 10.1038/nphys1856} {\bibfield
  {journal} {\bibinfo  {journal} {Nat. Phys.}\ }\textbf {\bibinfo {volume}
  {7}},\ \bibinfo {pages} {109} (\bibinfo {year} {2011})}\BibitemShut {NoStop}%
\bibitem [{\citenamefont {Malinowski}\ \emph {et~al.}(2016)\citenamefont
  {Malinowski}, \citenamefont {Martins}, \citenamefont {Nissen}, \citenamefont
  {Barnes}, \citenamefont {Rudner}, \citenamefont {Fallahi}, \citenamefont
  {Gardner}, \citenamefont {Manfra}, \citenamefont {Marcus},\ and\
  \citenamefont {Kuemmeth}}]{Malinowski2016}%
  \BibitemOpen
  \bibfield  {author} {\bibinfo {author} {\bibfnamefont {F.~K.}\ \bibnamefont
  {Malinowski}}, \bibinfo {author} {\bibfnamefont {F.}~\bibnamefont {Martins}},
  \bibinfo {author} {\bibfnamefont {P.~D.}\ \bibnamefont {Nissen}}, \bibinfo
  {author} {\bibfnamefont {E.}~\bibnamefont {Barnes}}, \bibinfo
  {author} {\bibfnamefont {\L{}.}~\bibnamefont {Cywi\'nski}}, \bibinfo {author}
  {\bibfnamefont {M.~S.}\ \bibnamefont {Rudner}}, \bibinfo {author}
  {\bibfnamefont {S.}~\bibnamefont {Fallahi}}, \bibinfo {author} {\bibfnamefont
  {G.~C.}\ \bibnamefont {Gardner}}, \bibinfo {author} {\bibfnamefont {M.~J.}\
  \bibnamefont {Manfra}}, \bibinfo {author} {\bibfnamefont {C.~M.}\
  \bibnamefont {Marcus}}, \ and\ \bibinfo {author} {\bibfnamefont
  {F.}~\bibnamefont {Kuemmeth}},\ }\href {\doibase 10.1038/nnano.2016.170}
  {\bibfield  {journal} {\bibinfo  {journal} {Nat. Nano.}\ }\textbf {\bibinfo
  {volume} {12}},\ \bibinfo {pages} {16} (\bibinfo {year} {2016})}\BibitemShut
  {NoStop}%
\bibitem [{\citenamefont {Rohling}\ and\ \citenamefont
  {Burkard}(2016)}]{Rohling2016}%
  \BibitemOpen
  \bibfield  {author} {\bibinfo {author} {\bibfnamefont {N.}~\bibnamefont
  {Rohling}}\ and\ \bibinfo {author} {\bibfnamefont {G.}~\bibnamefont
  {Burkard}},\ }\href {\doibase 10.1103/PhysRevB.93.205434} {\bibfield
  {journal} {\bibinfo  {journal} {Phys. Rev. B}\ }\textbf {\bibinfo {volume}
  {93}},\ \bibinfo {pages} {205434} (\bibinfo {year} {2016})}\BibitemShut
  {NoStop}%
\bibitem [{\citenamefont {Shulman}\ \emph {et~al.}(2014)\citenamefont
  {Shulman}, \citenamefont {Harvey}, \citenamefont {Nichol}, \citenamefont
  {Bartlett}, \citenamefont {Doherty}, \citenamefont {Umansky},\ and\
  \citenamefont {Yacoby}}]{Shulman2014}%
  \BibitemOpen
  \bibfield  {author} {\bibinfo {author} {\bibfnamefont {M.~D.}\ \bibnamefont
  {Shulman}}, \bibinfo {author} {\bibfnamefont {S.~P.}\ \bibnamefont {Harvey}},
  \bibinfo {author} {\bibfnamefont {J.~M.}\ \bibnamefont {Nichol}}, \bibinfo
  {author} {\bibfnamefont {S.~D.}\ \bibnamefont {Bartlett}}, \bibinfo {author}
  {\bibfnamefont {A.~C.}\ \bibnamefont {Doherty}}, \bibinfo {author}
  {\bibfnamefont {V.}~\bibnamefont {Umansky}}, \ and\ \bibinfo {author}
  {\bibfnamefont {A.}~\bibnamefont {Yacoby}},\ }\href
  {http://dx.doi.org/10.1038/ncomms6156} {\bibfield  {journal} {\bibinfo
  {journal} {Nature Communications}\ }\textbf {\bibinfo {volume} {5}},\
  \bibinfo {pages} {5156} (\bibinfo {year} {2014})}\BibitemShut {NoStop}%
\bibitem [{\citenamefont {Zwanenburg}\ \emph {et~al.}(2013)\citenamefont
  {Zwanenburg}, \citenamefont {Dzurak}, \citenamefont {Morello}, \citenamefont
  {Simmons}, \citenamefont {Hollenberg}, \citenamefont {Klimeck}, \citenamefont
  {Rogge}, \citenamefont {Coppersmith},\ and\ \citenamefont
  {Eriksson}}]{Zwanenburg}%
  \BibitemOpen
  \bibfield  {author} {\bibinfo {author} {\bibfnamefont {F.~A.}\ \bibnamefont
  {Zwanenburg}}, \bibinfo {author} {\bibfnamefont {A.~S.}\ \bibnamefont
  {Dzurak}}, \bibinfo {author} {\bibfnamefont {A.}~\bibnamefont {Morello}},
  \bibinfo {author} {\bibfnamefont {M.~Y.}\ \bibnamefont {Simmons}}, \bibinfo
  {author} {\bibfnamefont {L.~C.~L.}\ \bibnamefont {Hollenberg}}, \bibinfo
  {author} {\bibfnamefont {G.}~\bibnamefont {Klimeck}}, \bibinfo {author}
  {\bibfnamefont {S.}~\bibnamefont {Rogge}}, \bibinfo {author} {\bibfnamefont
  {S.~N.}\ \bibnamefont {Coppersmith}}, \ and\ \bibinfo {author} {\bibfnamefont
  {M.~A.}\ \bibnamefont {Eriksson}},\ }\href {\doibase
  10.1103/RevModPhys.85.961} {\bibfield  {journal} {\bibinfo  {journal} {Rev.
  Mod. Phys.}\ }\textbf {\bibinfo {volume} {85}},\ \bibinfo {pages} {961}
  (\bibinfo {year} {2013})}\BibitemShut {NoStop}%
\bibitem [{\citenamefont {Eng}\ \emph {et~al.}(2015)\citenamefont {Eng},
  \citenamefont {Ladd}, \citenamefont {Smith}, \citenamefont {Borselli},
  \citenamefont {Kiselev}, \citenamefont {Fong}, \citenamefont {Holabird},
  \citenamefont {Hazard}, \citenamefont {Huang}, \citenamefont {Deelman},
  \citenamefont {Milosavljevic}, \citenamefont {Schmitz}, \citenamefont {Ross},
  \citenamefont {Gyure},\ and\ \citenamefont {Hunter}}]{Eng}%
  \BibitemOpen
  \bibfield  {author} {\bibinfo {author} {\bibfnamefont {K.}~\bibnamefont
  {Eng}}, \bibinfo {author} {\bibfnamefont {T.~D.}\ \bibnamefont {Ladd}},
  \bibinfo {author} {\bibfnamefont {A.}~\bibnamefont {Smith}}, \bibinfo
  {author} {\bibfnamefont {M.~G.}\ \bibnamefont {Borselli}}, \bibinfo {author}
  {\bibfnamefont {A.~A.}\ \bibnamefont {Kiselev}}, \bibinfo {author}
  {\bibfnamefont {B.~H.}\ \bibnamefont {Fong}}, \bibinfo {author}
  {\bibfnamefont {K.~S.}\ \bibnamefont {Holabird}}, \bibinfo {author}
  {\bibfnamefont {T.~M.}\ \bibnamefont {Hazard}}, \bibinfo {author}
  {\bibfnamefont {B.}~\bibnamefont {Huang}}, \bibinfo {author} {\bibfnamefont
  {P.~W.}\ \bibnamefont {Deelman}}, \bibinfo {author} {\bibfnamefont
  {I.}~\bibnamefont {Milosavljevic}}, \bibinfo {author} {\bibfnamefont {A.~E.}\
  \bibnamefont {Schmitz}}, \bibinfo {author} {\bibfnamefont {R.~S.}\
  \bibnamefont {Ross}}, \bibinfo {author} {\bibfnamefont {M.~F.}\ \bibnamefont
  {Gyure}}, \ and\ \bibinfo {author} {\bibfnamefont {A.~T.}\ \bibnamefont
  {Hunter}},\ }\href {\doibase 10.1126/sciadv.1500214} {\bibfield  {journal}
  {\bibinfo  {journal} {Science Adv.}\ }\textbf {\bibinfo {volume} {1}},\ \bibinfo {pages}
  {e1500214} (\bibinfo {year} {2015})}\BibitemShut {NoStop}%
\bibitem [{\citenamefont {Veldhorst}\ \emph {et~al.}(2014)\citenamefont
  {Veldhorst}, \citenamefont {Hwang}, \citenamefont {Yang}, \citenamefont
  {Leenstra}, \citenamefont {de~Ronde}, \citenamefont {Dehollain},
  \citenamefont {Muhonen}, \citenamefont {Hudson}, \citenamefont {Itoh},
  \citenamefont {Morello},\ and\ \citenamefont {Dzurak}}]{Veldhorst2014}%
  \BibitemOpen
  \bibfield  {author} {\bibinfo {author} {\bibfnamefont {M.}~\bibnamefont
  {Veldhorst}}, \bibinfo {author} {\bibfnamefont {J.~C.~C.}\ \bibnamefont
  {Hwang}}, \bibinfo {author} {\bibfnamefont {C.~H.}\ \bibnamefont {Yang}},
  \bibinfo {author} {\bibfnamefont {A.~W.}\ \bibnamefont {Leenstra}}, \bibinfo
  {author} {\bibfnamefont {B.}~\bibnamefont {de~Ronde}}, \bibinfo {author}
  {\bibfnamefont {J.~P.}\ \bibnamefont {Dehollain}}, \bibinfo {author}
  {\bibfnamefont {J.~T.}\ \bibnamefont {Muhonen}}, \bibinfo {author}
  {\bibfnamefont {F.~E.}\ \bibnamefont {Hudson}}, \bibinfo {author}
  {\bibfnamefont {K.~M.}\ \bibnamefont {Itoh}}, \bibinfo {author}
  {\bibfnamefont {A.}~\bibnamefont {Morello}}, \ and\ \bibinfo {author}
  {\bibfnamefont {A.~S.}\ \bibnamefont {Dzurak}},\ }\href
  {http://dx.doi.org/10.1038/nnano.2014.216} {\bibfield  {journal} {\bibinfo
  {journal} {Nat Nano}\ }\textbf {\bibinfo {volume} {9}},\ \bibinfo {pages}
  {981} (\bibinfo {year} {2014})}\BibitemShut
  {NoStop}%
\bibitem [{\citenamefont {Bacon}\ \emph {et~al.}(2000)\citenamefont {Bacon},
  \citenamefont {Kempe}, \citenamefont {Lidar},\ and\ \citenamefont
  {Whaley}}]{Bacon2000}%
  \BibitemOpen
  \bibfield  {author} {\bibinfo {author} {\bibfnamefont {D.}~\bibnamefont
  {Bacon}}, \bibinfo {author} {\bibfnamefont {J.}~\bibnamefont {Kempe}},
  \bibinfo {author} {\bibfnamefont {D.~A.}\ \bibnamefont {Lidar}}, \ and\
  \bibinfo {author} {\bibfnamefont {K.~B.}\ \bibnamefont {Whaley}},\ }\href
  {\doibase 10.1103/PhysRevLett.85.1758} {\bibfield  {journal} {\bibinfo
  {journal} {Phys. Rev. Lett.}\ }\textbf {\bibinfo {volume} {85}},\ \bibinfo
  {pages} {1758} (\bibinfo {year} {2000})}\BibitemShut {NoStop}%
\bibitem [{\citenamefont {Lidar}\ and\ \citenamefont {{Birgitta
  Whaley}}(2003)}]{Lidar2003}%
  \BibitemOpen
  \bibfield  {author} {\bibinfo {author} {\bibfnamefont {D.~A.}\ \bibnamefont
  {Lidar}}\ and\ \bibinfo {author} {\bibfnamefont {K.}~\bibnamefont {{Birgitta
  Whaley}}},\ }in\ \href {\doibase 10.1007/3-540-44874-8_5} {\emph {\bibinfo
  {booktitle} {Irreversible Quantum Dynamics}}},\ \bibinfo {editor} {edited by\
  \bibinfo {editor} {\bibfnamefont {F.}~\bibnamefont {Benatti}}\ and\ \bibinfo
  {editor} {\bibfnamefont {R.}~\bibnamefont {Floreanini}}}\ (\bibinfo
  {publisher} {Springer-Verlag Berlin Heidelberg},\ \bibinfo {year} {2003})\
  pp.\ \bibinfo {pages} {83--120}\BibitemShut {NoStop}%
\bibitem [{\citenamefont {Scarola}\ \emph {et~al.}(2004)\citenamefont
  {Scarola}, \citenamefont {Park},\ and\ \citenamefont {{Das
  Sarma}}}]{Scarola2004}%
  \BibitemOpen
  \bibfield  {author} {\bibinfo {author} {\bibfnamefont {V.~W.}\ \bibnamefont
  {Scarola}}, \bibinfo {author} {\bibfnamefont {K.}~\bibnamefont {Park}}, \
  and\ \bibinfo {author} {\bibfnamefont {S.}~\bibnamefont {{Das Sarma}}},\
  }\href {\doibase 10.1103/PhysRevLett.93.120503} {\bibfield  {journal}
  {\bibinfo  {journal} {Phys. Rev. Lett.}\ }\textbf {\bibinfo {volume} {93}},\
  \bibinfo {pages} {120503} (\bibinfo {year} {2004})}\BibitemShut {NoStop}%
\bibitem [{\citenamefont {Antonio}\ and\ \citenamefont
  {Bose}(2013)}]{Antonio2013}%
  \BibitemOpen
  \bibfield  {author} {\bibinfo {author} {\bibfnamefont {B.}~\bibnamefont
  {Antonio}}\ and\ \bibinfo {author} {\bibfnamefont {S.}~\bibnamefont {Bose}},\
  }\href {\doibase 10.1103/PhysRevA.88.042306} {\bibfield  {journal} {\bibinfo
  {journal} {Phys. Rev. A}\ }\textbf {\bibinfo {volume} {88}},\ \bibinfo
  {pages} {042306} (\bibinfo {year} {2013})}\BibitemShut {NoStop}%
\bibitem [{Note1()}]{Note1}%
  \BibitemOpen
  \bibinfo {note} {In fact, if sensitive enough, one charge sensor could
  suffice, as long as it is not placed equidistant from any pair of
  dots.}\BibitemShut {Stop}%
\bibitem [{\citenamefont {Burkard}\ \emph {et~al.}(1999)\citenamefont
  {Burkard}, \citenamefont {Loss},\ and\ \citenamefont
  {DiVincenzo}}]{burkard_prb}%
  \BibitemOpen
  \bibfield  {author} {\bibinfo {author} {\bibfnamefont {G.}~\bibnamefont
  {Burkard}}, \bibinfo {author} {\bibfnamefont {D.}~\bibnamefont {Loss}}, \
  and\ \bibinfo {author} {\bibfnamefont {D.~P.}\ \bibnamefont {DiVincenzo}},\
  }\href {\doibase 10.1103/PhysRevB.59.2070} {\bibfield  {journal} {\bibinfo
  {journal} {Phys. Rev. B}\ }\textbf {\bibinfo {volume} {59}},\ \bibinfo
  {pages} {2070} (\bibinfo {year} {1999})}\BibitemShut {NoStop}%
\bibitem [{\citenamefont {Taylor}\ \emph {et~al.}(2013)\citenamefont {Taylor},
  \citenamefont {Srinivasa},\ and\ \citenamefont {Medford}}]{Taylor}%
  \BibitemOpen
  \bibfield  {author} {\bibinfo {author} {\bibfnamefont {J.~M.}\ \bibnamefont
  {Taylor}}, \bibinfo {author} {\bibfnamefont {V.}~\bibnamefont {Srinivasa}}, \
  and\ \bibinfo {author} {\bibfnamefont {J.}~\bibnamefont {Medford}},\ }\href
  {\doibase 10.1103/PhysRevLett.111.050502} {\bibfield  {journal} {\bibinfo
  {journal} {Phys. Rev. Lett.}\ }\textbf {\bibinfo {volume} {111}},\ \bibinfo
  {pages} {050502} (\bibinfo {year} {2013})}\BibitemShut {NoStop}%
\bibitem [{Note2()}]{Note2}%
  \BibitemOpen
  \bibinfo {note} {See the Supplementary Material for more detailed discussions
  and derivations of the equations presented in the main text, which includes 
  Refs.~\cite{Schrieffer1966,Danon2013,Thalineau2012,Takakura2014a,Delbecq2014,Zajac2016}.}\BibitemShut
  {Stop}%
\bibitem [{\citenamefont {Schrieffer}\ and\ \citenamefont
  {Wolff}(1966)}]{Schrieffer1966}%
  \BibitemOpen
  \bibfield  {author} {\bibinfo {author} {\bibfnamefont {J.~R.}\ \bibnamefont
  {Schrieffer}}\ and\ \bibinfo {author} {\bibfnamefont {P.~A.}\ \bibnamefont
  {Wolff}},\ }\href {\doibase 10.1103/PhysRev.149.491} {\bibfield  {journal}
  {\bibinfo  {journal} {Phys. Rev.}\ }\textbf {\bibinfo {volume} {149}},\
  \bibinfo {pages} {491} (\bibinfo {year} {1966})}\BibitemShut {NoStop}%
\bibitem [{\citenamefont {Danon}(2013)}]{Danon2013}%
  \BibitemOpen
  \bibfield  {author} {\bibinfo {author} {\bibfnamefont {J.}~\bibnamefont
  {Danon}},\ }\href {\doibase 10.1103/PhysRevB.88.075306} {\bibfield  {journal}
  {\bibinfo  {journal} {Phys. Rev. B}\ }\textbf {\bibinfo {volume} {88}},\
  \bibinfo {pages} {075306} (\bibinfo {year} {2013})}\BibitemShut {NoStop}%
\bibitem [{\citenamefont {Thalineau}\ \emph {et~al.}(2012)\citenamefont
  {Thalineau}, \citenamefont {Hermelin}, \citenamefont {Wieck}, \citenamefont
  {B{\"{a}}uerle}, \citenamefont {Saminadayar},\ and\ \citenamefont
  {Meunier}}]{Thalineau2012}%
  \BibitemOpen
  \bibfield  {author} {\bibinfo {author} {\bibfnamefont {R.}~\bibnamefont
  {Thalineau}}, \bibinfo {author} {\bibfnamefont {S.}~\bibnamefont {Hermelin}},
  \bibinfo {author} {\bibfnamefont {A.~D.}\ \bibnamefont {Wieck}}, \bibinfo
  {author} {\bibfnamefont {C.}~\bibnamefont {B{\"{a}}uerle}}, \bibinfo {author}
  {\bibfnamefont {L.}~\bibnamefont {Saminadayar}}, \ and\ \bibinfo {author}
  {\bibfnamefont {T.}~\bibnamefont {Meunier}},\ }\href
  {http://aip.scitation.org/doi/10.1063/1.4749811} {\bibfield  {journal}
  {\bibinfo  {journal} {Appl. Phys. Lett.}\ }\textbf {\bibinfo {volume}
  {101}},\ \bibinfo {pages} {103102} (\bibinfo {year} {2012})}\BibitemShut
  {NoStop}%
\bibitem [{\citenamefont {Takakura}\ \emph {et~al.}(2014)\citenamefont
  {Takakura}, \citenamefont {Noiri}, \citenamefont {Obata}, \citenamefont
  {Otsuka}, \citenamefont {Yoneda}, \citenamefont {Yoshida},\ and\
  \citenamefont {Tarucha}}]{Takakura2014a}%
  \BibitemOpen
  \bibfield  {author} {\bibinfo {author} {\bibfnamefont {T.}~\bibnamefont
  {Takakura}}, \bibinfo {author} {\bibfnamefont {A.}~\bibnamefont {Noiri}},
  \bibinfo {author} {\bibfnamefont {T.}~\bibnamefont {Obata}}, \bibinfo
  {author} {\bibfnamefont {T.}~\bibnamefont {Otsuka}}, \bibinfo {author}
  {\bibfnamefont {J.}~\bibnamefont {Yoneda}}, \bibinfo {author} {\bibfnamefont
  {K.}~\bibnamefont {Yoshida}}, \ and\ \bibinfo {author} {\bibfnamefont
  {S.}~\bibnamefont {Tarucha}},\ }\href
  {http://aip.scitation.org/doi/10.1063/1.4869108} {\bibfield  {journal}
  {\bibinfo  {journal} {Appl. Phys. Lett.}\ }\textbf {\bibinfo {volume}
  {104}},\ \bibinfo {pages} {113109} (\bibinfo {year} {2014})}\BibitemShut
  {NoStop}%
\bibitem [{\citenamefont {Delbecq}\ \emph {et~al.}(2014)\citenamefont
  {Delbecq}, \citenamefont {Nakajima}, \citenamefont {Otsuka}, \citenamefont
  {Amaha}, \citenamefont {Watson}, \citenamefont {Manfra},\ and\ \citenamefont
  {Tarucha}}]{Delbecq2014}%
  \BibitemOpen
  \bibfield  {author} {\bibinfo {author} {\bibfnamefont {M.~R.}\ \bibnamefont
  {Delbecq}}, \bibinfo {author} {\bibfnamefont {T.}~\bibnamefont {Nakajima}},
  \bibinfo {author} {\bibfnamefont {T.}~\bibnamefont {Otsuka}}, \bibinfo
  {author} {\bibfnamefont {S.}~\bibnamefont {Amaha}}, \bibinfo {author}
  {\bibfnamefont {J.~D.}\ \bibnamefont {Watson}}, \bibinfo {author}
  {\bibfnamefont {M.~J.}\ \bibnamefont {Manfra}}, \ and\ \bibinfo {author}
  {\bibfnamefont {S.}~\bibnamefont {Tarucha}},\ }\href
  {http://aip.scitation.org/doi/10.1063/1.4875909} {\bibfield  {journal}
  {\bibinfo  {journal} {Appl. Phys. Lett.}\ }\textbf {\bibinfo {volume}
  {104}},\ \bibinfo {pages} {183111} (\bibinfo {year} {2014})}\BibitemShut
  {NoStop}%
\bibitem [{\citenamefont {Zajac}\ \emph {et~al.}(2016)\citenamefont {Zajac},
  \citenamefont {Hazard}, \citenamefont {Mi}, \citenamefont {Nielsen},\ and\
  \citenamefont {Petta}}]{Zajac2016}%
  \BibitemOpen
  \bibfield  {author} {\bibinfo {author} {\bibfnamefont {D.~M.}\ \bibnamefont
  {Zajac}}, \bibinfo {author} {\bibfnamefont {T.~M.}\ \bibnamefont {Hazard}},
  \bibinfo {author} {\bibfnamefont {X.}~\bibnamefont {Mi}}, \bibinfo {author}
  {\bibfnamefont {E.}~\bibnamefont {Nielsen}}, \ and\ \bibinfo {author}
  {\bibfnamefont {J.~R.}\ \bibnamefont {Petta}},\ }\href {\doibase
  10.1103/PhysRevApplied.6.054013} {\bibfield  {journal} {\bibinfo  {journal}
  {Phys. Rev. Appl.}\ }\textbf {\bibinfo {volume} {6}},\ \bibinfo {pages}
  {054013} (\bibinfo {year} {2016})}\BibitemShut {NoStop}%
\bibitem [{\citenamefont {Johnson}\ \emph {et~al.}(2005)\citenamefont
  {Johnson}, \citenamefont {Petta}, \citenamefont {Taylor}, \citenamefont
  {Yacoby}, \citenamefont {Lukin}, \citenamefont {Marcus}, \citenamefont
  {Hanson},\ and\ \citenamefont {Gossard}}]{Johnson}%
  \BibitemOpen
  \bibfield  {author} {\bibinfo {author} {\bibfnamefont {A.~C.}\ \bibnamefont
  {Johnson}}, \bibinfo {author} {\bibfnamefont {J.~R.}\ \bibnamefont {Petta}},
  \bibinfo {author} {\bibfnamefont {J.~M.}\ \bibnamefont {Taylor}}, \bibinfo
  {author} {\bibfnamefont {A.}~\bibnamefont {Yacoby}}, \bibinfo {author}
  {\bibfnamefont {M.~D.}\ \bibnamefont {Lukin}}, \bibinfo {author}
  {\bibfnamefont {C.~M.}\ \bibnamefont {Marcus}}, \bibinfo {author}
  {\bibfnamefont {M.~P.}\ \bibnamefont {Hanson}}, \ and\ \bibinfo {author}
  {\bibfnamefont {A.~C.}\ \bibnamefont {Gossard}},\ }\href {\doibase
  10.1038/nature03815} {\bibfield  {journal} {\bibinfo  {journal} {Nature}\
  }\textbf {\bibinfo {volume} {435}},\ \bibinfo {pages} {925} (\bibinfo {year}
  {2005})}\BibitemShut {NoStop}%
\bibitem [{Note3()}]{Note3}%
  \BibitemOpen
  \bibinfo {note} {$K_i^{x,y}$ can cause a similar higher-order shift in the
  qubit splitting, but the resulting contribution to the dephasing rate
  $1/T_2^*$ is smaller by a factor $J/E_{\protect \rm Z}$}\BibitemShut
  {NoStop}%
\bibitem [{Note4()}]{Note4}%
  \BibitemOpen
  \bibinfo {note} {We note that these simulations include hyperfine-induced
  leakage out of the qubit space to other states with $S_z=0$.}\BibitemShut
  {Stop}%
\bibitem [{\citenamefont {Stano}\ and\ \citenamefont
  {Fabian}(2006)}]{PhysRevB.74.045320}%
  \BibitemOpen
  \bibfield  {author} {\bibinfo {author} {\bibfnamefont {P.}~\bibnamefont
  {Stano}}\ and\ \bibinfo {author} {\bibfnamefont {J.}~\bibnamefont {Fabian}},\
  }\href {\doibase 10.1103/PhysRevB.74.045320} {\bibfield  {journal} {\bibinfo
  {journal} {Phys. Rev. B}\ }\textbf {\bibinfo {volume} {74}},\ \bibinfo
  {pages} {045320} (\bibinfo {year} {2006})}\BibitemShut {NoStop}%
\bibitem [{\citenamefont {Madelung}(2004)}]{Madelung2004}%
  \BibitemOpen
  \bibfield  {author} {\bibinfo {author} {\bibfnamefont {O.}~\bibnamefont
  {Madelung}},\ }\href {\doibase 10.1007/978-3-642-18865-7} {\emph {\bibinfo
  {title} {Semiconductors: Data Handbook}}},\ \bibinfo {edition} {3rd}\ ed.\
  (\bibinfo  {publisher} {Springer-Verlag Berlin Heidelberg},\ \bibinfo {year}
  {2004})\BibitemShut {NoStop}%
\bibitem [{Note5()}]{Note5}%
  \BibitemOpen
  \bibinfo {note} {Possible contributions from spin-orbit interaction to
  relaxation to $\ket {T_3}$ could be suppressed by applying the external
  magnetic field perpendicularly to the qubit plane.}\BibitemShut {Stop}%
\end{thebibliography}

\begin{thebibliography}{12}%
\makeatletter
\providecommand \@ifxundefined [1]{%
 \@ifx{#1\undefined}
}%
\providecommand \@ifnum [1]{%
 \ifnum #1\expandafter \@firstoftwo
 \else \expandafter \@secondoftwo
 \fi
}%
\providecommand \@ifx [1]{%
 \ifx #1\expandafter \@firstoftwo
 \else \expandafter \@secondoftwo
 \fi
}%
\providecommand \natexlab [1]{#1}%
\providecommand \enquote  [1]{``#1''}%
\providecommand \bibnamefont  [1]{#1}%
\providecommand \bibfnamefont [1]{#1}%
\providecommand \citenamefont [1]{#1}%
\providecommand \href@noop [0]{\@secondoftwo}%
\providecommand \href [0]{\begingroup \@sanitize@url \@href}%
\providecommand \@href[1]{\@@startlink{#1}\@@href}%
\providecommand \@@href[1]{\endgroup#1\@@endlink}%
\providecommand \@sanitize@url [0]{\catcode `\\12\catcode `\$12\catcode
  `\&12\catcode `\#12\catcode `\^12\catcode `\_12\catcode `\%12\relax}%
\providecommand \@@startlink[1]{}%
\providecommand \@@endlink[0]{}%
\providecommand \url  [0]{\begingroup\@sanitize@url \@url }%
\providecommand \@url [1]{\endgroup\@href {#1}{\urlprefix }}%
\providecommand \urlprefix  [0]{URL }%
\providecommand \Eprint [0]{\href }%
\providecommand \doibase [0]{http://dx.doi.org/}%
\providecommand \selectlanguage [0]{\@gobble}%
\providecommand \bibinfo  [0]{\@secondoftwo}%
\providecommand \bibfield  [0]{\@secondoftwo}%
\providecommand \translation [1]{[#1]}%
\providecommand \BibitemOpen [0]{}%
\providecommand \bibitemStop [0]{}%
\providecommand \bibitemNoStop [0]{.\EOS\space}%
\providecommand \EOS [0]{\spacefactor3000\relax}%
\providecommand \BibitemShut  [1]{\csname bibitem#1\endcsname}%
\let\auto@bib@innerbib\@empty
\bibitem [{\citenamefont {Burkard}\ \emph {et~al.}(1999)\citenamefont
  {Burkard}, \citenamefont {Loss},\ and\ \citenamefont
  {DiVincenzo}}]{SUPburkard_prb}%
  \BibitemOpen
  \bibfield  {author} {\bibinfo {author} {\bibfnamefont {G.}~\bibnamefont
  {Burkard}}, \bibinfo {author} {\bibfnamefont {D.}~\bibnamefont {Loss}}, \
  and\ \bibinfo {author} {\bibfnamefont {D.~P.}\ \bibnamefont {DiVincenzo}},\
  }\href {\doibase 10.1103/PhysRevB.59.2070} {\bibfield  {journal} {\bibinfo
  {journal} {Phys. Rev. B}\ }\textbf {\bibinfo {volume} {59}},\ \bibinfo
  {pages} {2070} (\bibinfo {year} {1999})}\BibitemShut {NoStop}%
\bibitem [{\citenamefont {Taylor}\ \emph {et~al.}(2013)\citenamefont {Taylor},
  \citenamefont {Srinivasa},\ and\ \citenamefont {Medford}}]{SUPTaylor}%
  \BibitemOpen
  \bibfield  {author} {\bibinfo {author} {\bibfnamefont {J.~M.}\ \bibnamefont
  {Taylor}}, \bibinfo {author} {\bibfnamefont {V.}~\bibnamefont {Srinivasa}}, \
  and\ \bibinfo {author} {\bibfnamefont {J.}~\bibnamefont {Medford}},\ }\href
  {\doibase 10.1103/PhysRevLett.111.050502} {\bibfield  {journal} {\bibinfo
  {journal} {Phys. Rev. Lett.}\ }\textbf {\bibinfo {volume} {111}},\ \bibinfo
  {pages} {050502} (\bibinfo {year} {2013})}\BibitemShut {NoStop}%
\bibitem [{\citenamefont {Schrieffer}\ and\ \citenamefont
  {Wolff}(1966)}]{SUPSchrieffer1966}%
  \BibitemOpen
  \bibfield  {author} {\bibinfo {author} {\bibfnamefont {J.~R.}\ \bibnamefont
  {Schrieffer}}\ and\ \bibinfo {author} {\bibfnamefont {P.~A.}\ \bibnamefont
  {Wolff}},\ }\href {\doibase 10.1103/PhysRev.149.491} {\bibfield  {journal}
  {\bibinfo  {journal} {Phys. Rev.}\ }\textbf {\bibinfo {volume} {149}},\
  \bibinfo {pages} {491} (\bibinfo {year} {1966})}\BibitemShut {NoStop}%
\bibitem [{Note1()}]{SUPNote1}%
  \BibitemOpen
  \bibinfo {note} {These states are now written to zeroth order in the
  $t_{ij}$. Below we will give the first-order corrections the states acquire
  due to the transformation.}\BibitemShut {Stop}%
\bibitem [{Note2()}]{SUPNote2}%
  \BibitemOpen
  \bibinfo {note} {In the main text we wrote the Hamiltonian as a function of
  the detuning parameter $\epsilon _{14}$ only, i.e., we wrote the Hamiltonian
  setting $\epsilon _{23} = \epsilon _\Lambda = 0$.}\BibitemShut {Stop}%
\bibitem [{\citenamefont {Danon}(2013)}]{SUPDanon2013}%
  \BibitemOpen
  \bibfield  {author} {\bibinfo {author} {\bibfnamefont {J.}~\bibnamefont
  {Danon}},\ }\href {\doibase 10.1103/PhysRevB.88.075306} {\bibfield  {journal}
  {\bibinfo  {journal} {Phys. Rev. B}\ }\textbf {\bibinfo {volume} {88}},\
  \bibinfo {pages} {075306} (\bibinfo {year} {2013})}\BibitemShut {NoStop}%
\bibitem [{Note3()}]{SUPNote3}%
  \BibitemOpen
  \bibinfo {note} {Note that the Hamiltonian~(\ref {eq:shamt}) already takes
  into account the finite overlap between electronic wave functions on
  neighboring dots.}\BibitemShut {Stop}%
\bibitem [{\citenamefont {Medford}\ \emph {et~al.}(2013)\citenamefont
  {Medford}, \citenamefont {Beil}, \citenamefont {Taylor}, \citenamefont
  {Bartlett}, \citenamefont {Doherty}, \citenamefont {Rashba}, \citenamefont
  {DiVincenzo}, \citenamefont {Lu}, \citenamefont {Gossard},\ and\
  \citenamefont {Marcus}}]{SUPMedford}%
  \BibitemOpen
  \bibfield  {author} {\bibinfo {author} {\bibfnamefont {J.}~\bibnamefont
  {Medford}}, \bibinfo {author} {\bibfnamefont {J.}~\bibnamefont {Beil}},
  \bibinfo {author} {\bibfnamefont {J.~M.}\ \bibnamefont {Taylor}}, \bibinfo
  {author} {\bibfnamefont {S.~D.}\ \bibnamefont {Bartlett}}, \bibinfo {author}
  {\bibfnamefont {A.~C.}\ \bibnamefont {Doherty}}, \bibinfo {author}
  {\bibfnamefont {E.~I.}\ \bibnamefont {Rashba}}, \bibinfo {author}
  {\bibfnamefont {D.~P.}\ \bibnamefont {DiVincenzo}}, \bibinfo {author}
  {\bibfnamefont {H.}~\bibnamefont {Lu}}, \bibinfo {author} {\bibfnamefont
  {A.~C.}\ \bibnamefont {Gossard}}, \ and\ \bibinfo {author} {\bibfnamefont
  {C.~M.}\ \bibnamefont {Marcus}},\ }\href
  {http://dx.doi.org/10.1038/nnano.2013.168} {\bibfield  {journal} {\bibinfo
  {journal} {Nat. Nano.}\ }\textbf {\bibinfo {volume} {8}},\ \bibinfo {pages}
  {654} (\bibinfo {year} {2013})}\BibitemShut {NoStop}%
\bibitem [{\citenamefont {Thalineau}\ \emph {et~al.}(2012)\citenamefont
  {Thalineau}, \citenamefont {Hermelin}, \citenamefont {Wieck}, \citenamefont
  {B{\"{a}}uerle}, \citenamefont {Saminadayar},\ and\ \citenamefont
  {Meunier}}]{SUPThalineau2012}%
  \BibitemOpen
  \bibfield  {author} {\bibinfo {author} {\bibfnamefont {R.}~\bibnamefont
  {Thalineau}}, \bibinfo {author} {\bibfnamefont {S.}~\bibnamefont {Hermelin}},
  \bibinfo {author} {\bibfnamefont {A.~D.}\ \bibnamefont {Wieck}}, \bibinfo
  {author} {\bibfnamefont {C.}~\bibnamefont {B{\"{a}}uerle}}, \bibinfo {author}
  {\bibfnamefont {L.}~\bibnamefont {Saminadayar}}, \ and\ \bibinfo {author}
  {\bibfnamefont {T.}~\bibnamefont {Meunier}},\ }\href
  {http://aip.scitation.org/doi/10.1063/1.4749811} {\bibfield  {journal}
  {\bibinfo  {journal} {Appl. Phys. Lett.}\ }\textbf {\bibinfo {volume}
  {101}},\ \bibinfo {pages} {103102} (\bibinfo {year} {2012})}\BibitemShut
  {NoStop}%
\bibitem [{\citenamefont {Takakura}\ \emph {et~al.}(2014)\citenamefont
  {Takakura}, \citenamefont {Noiri}, \citenamefont {Obata}, \citenamefont
  {Otsuka}, \citenamefont {Yoneda}, \citenamefont {Yoshida},\ and\
  \citenamefont {Tarucha}}]{SUPTakakura2014a}%
  \BibitemOpen
  \bibfield  {author} {\bibinfo {author} {\bibfnamefont {T.}~\bibnamefont
  {Takakura}}, \bibinfo {author} {\bibfnamefont {A.}~\bibnamefont {Noiri}},
  \bibinfo {author} {\bibfnamefont {T.}~\bibnamefont {Obata}}, \bibinfo
  {author} {\bibfnamefont {T.}~\bibnamefont {Otsuka}}, \bibinfo {author}
  {\bibfnamefont {J.}~\bibnamefont {Yoneda}}, \bibinfo {author} {\bibfnamefont
  {K.}~\bibnamefont {Yoshida}}, \ and\ \bibinfo {author} {\bibfnamefont
  {S.}~\bibnamefont {Tarucha}},\ }\href
  {http://aip.scitation.org/doi/10.1063/1.4869108} {\bibfield  {journal}
  {\bibinfo  {journal} {Appl. Phys. Lett.}\ }\textbf {\bibinfo {volume}
  {104}},\ \bibinfo {pages} {113109} (\bibinfo {year} {2014})}\BibitemShut
  {NoStop}%
\bibitem [{\citenamefont {Delbecq}\ \emph {et~al.}(2014)\citenamefont
  {Delbecq}, \citenamefont {Nakajima}, \citenamefont {Otsuka}, \citenamefont
  {Amaha}, \citenamefont {Watson}, \citenamefont {Manfra},\ and\ \citenamefont
  {Tarucha}}]{SUPDelbecq2014}%
  \BibitemOpen
  \bibfield  {author} {\bibinfo {author} {\bibfnamefont {M.~R.}\ \bibnamefont
  {Delbecq}}, \bibinfo {author} {\bibfnamefont {T.}~\bibnamefont {Nakajima}},
  \bibinfo {author} {\bibfnamefont {T.}~\bibnamefont {Otsuka}}, \bibinfo
  {author} {\bibfnamefont {S.}~\bibnamefont {Amaha}}, \bibinfo {author}
  {\bibfnamefont {J.~D.}\ \bibnamefont {Watson}}, \bibinfo {author}
  {\bibfnamefont {M.~J.}\ \bibnamefont {Manfra}}, \ and\ \bibinfo {author}
  {\bibfnamefont {S.}~\bibnamefont {Tarucha}},\ }\href
  {http://aip.scitation.org/doi/10.1063/1.4875909} {\bibfield  {journal}
  {\bibinfo  {journal} {Appl. Phys. Lett.}\ }\textbf {\bibinfo {volume}
  {104}},\ \bibinfo {pages} {183111} (\bibinfo {year} {2014})}\BibitemShut
  {NoStop}%
\bibitem [{\citenamefont {Zajac}\ \emph {et~al.}(2016)\citenamefont {Zajac},
  \citenamefont {Hazard}, \citenamefont {Mi}, \citenamefont {Nielsen},\ and\
  \citenamefont {Petta}}]{SUPZajac2016}%
  \BibitemOpen
  \bibfield  {author} {\bibinfo {author} {\bibfnamefont {D.~M.}\ \bibnamefont
  {Zajac}}, \bibinfo {author} {\bibfnamefont {T.~M.}\ \bibnamefont {Hazard}},
  \bibinfo {author} {\bibfnamefont {X.}~\bibnamefont {Mi}}, \bibinfo {author}
  {\bibfnamefont {E.}~\bibnamefont {Nielsen}}, \ and\ \bibinfo {author}
  {\bibfnamefont {J.~R.}\ \bibnamefont {Petta}},\ }\href {\doibase
  10.1103/PhysRevApplied.6.054013} {\bibfield  {journal} {\bibinfo  {journal}
  {Phys. Rev. Appl.}\ }\textbf {\bibinfo {volume} {6}},\ \bibinfo {pages}
  {054013} (\bibinfo {year} {2016})}\BibitemShut {NoStop}%
\end{thebibliography}
\end{document}